%% file: Main.tex
\newif\ifdoubleblind
\doubleblindfalse
\ifdoubleblind
\documentclass[manuscript,review,anonymous]{acmart}
\else
\documentclass[manuscript]{acmart}
\fi
\AtBeginDocument{%
  }

\usepackage{tikz}
\usepackage{amsmath}
\usepackage{xurl} %
\usepackage{natbib}
\usepackage{filecontents}

\usepackage{booktabs}
\usepackage{multirow}
\usepackage{cleveref}
\usepackage{enumitem}

\usepackage{array}
\usepackage{xcolor, colortbl}
\usepackage{geometry}
\usepackage{pdflscape}
\usepackage{caption}
\usepackage{tabularx}
\definecolor{colMM}{HTML}{EEEDFE}
\definecolor{colMTA}{HTML}{E6F1FB}
\definecolor{colCryptocurrency}{HTML}{E1F5EE}
\definecolor{colStable}{HTML}{FFF3E0}
\definecolor{colCBDC}{HTML}{FCE4EC}
\definecolor{headerMM}{HTML}{3C3489}
\definecolor{headerMTA}{HTML}{185FA5}
\definecolor{headerCryptocurrency}{HTML}{0F6E56}
\definecolor{headerStable}{HTML}{854F0B}
\definecolor{headerCBDC}{HTML}{993556}
\definecolor{rowlabel}{HTML}{F1EFE8}
\newcolumntype{P}[1]{>{\raggedright\arraybackslash}p{#1}}

\setcopyright{acmlicensed} %% CCS: DO NOT REMOVE
\copyrightyear{2018} %% CCS: DO NOT REMOVE
\acmYear{2018} %% CCS: DO NOT REMOVE
\acmDOI{XXXXXXX.XXXXXXX} %% CCS: DO NOT REMOVE
\acmConference[Conference acronym 'XX]{Make sure to enter the correct
  conference title from your rights confirmation email}{June 03--05,
  2018}{Woodstock, NY}  %% CCS: DO NOT REMOVE
\acmISBN{978-1-4503-XXXX-X/2018/06}  %% CCS: DO NOT REMOVE

\begin{document}
%-------------------------------------------------------------------------------

%don't want date printed
\date{}

\title{A Cross-Country Evaluation of Sentiment Toward Digital Payment Systems in Africa}

\newif\ifdraft
\drafttrue
% \draftfalse % uncomment to remove all comments
\ifdraft
    \newcommand{\gf}[1]{\textcolor{red}{GF: #1}}
    \newcommand{\nc}[1]{\textcolor{blue}{NC: #1}}
    \newcommand{\nicolasc}[1]{\textcolor{blue}{NC: #1}}
    \newcommand{\ks}[1]{\textcolor{cyan}{KS: #1}}
    \newcommand{\tk}[1]{\textcolor{orange}{TK: #1}}
    \newcommand{\ts}[1]{\textcolor{teal}{TS: #1}}
    \newcommand{\ia}[1]{\textcolor{purple}{IA: #1}}
\else
    \newcommand{\gf}[1]{}
    \newcommand{\nc}[1]{}
    \newcommand{\nicolasc}[1]{}
    \newcommand{\ks}[1]{}
    \newcommand{\tk}[1]{}
    \newcommand{\ts\}[1]{}
    \newcommand{\ia}[1]{}
\fi

%
% The "author" command and its associated commands are used to define
% the authors and their affiliations.
% Of note is the shared affiliation of the first two authors, and the
% "authornote" and "authornotemark" commands
% used to denote shared contribution to the research.

% CCS: at submission time, the submission MUST be anonymized. Hence
% authors MUST be commented out.

\ifdoubleblind
% \author{Anonymous Submission}
\else
\author{Isabel Agadagba}
\authornote{Authors contributed equally to this research.}
\email{iagadagb@alumni.cmu.edu}
\affiliation{%
  \institution{Carnegie Mellon University}
  % \city{Pittsburgh}
  % \state{Pennsylvania}
  \country{USA}
}

\author{Triphonia Kilasara}
\authornotemark[1]
\email{tkilasar@alumni.cmu.edu}
\affiliation{%
  \institution{Carnegie Mellon University-Africa}
  % \city{Pittsburgh}
  % \state{Pennsylvania}
  \country{Rwanda}
}

\author{Takudzwa Talent Tarutira}
\authornotemark[1]
\email{ttarutir@alumni.cmu.edu}
\affiliation{%
  \institution{Carnegie Mellon University}
  % \city{Pittsburgh}
  % \state{Pennsylvania}
  \country{USA}
}

\author{Noah Shumba}
\email{nshumba@andrew.cmu.edu }
\affiliation{%
  \institution{Carnegie Mellon University-Africa}
  % \city{Pittsburgh}
  % \state{Pennsylvania}
  \country{Rwanda}
}

\author{Nicolas Christin}
\email{nicolasc@andrew.cmu.edu}
\affiliation{%
  \institution{Carnegie Mellon University}
  % \city{Pittsburgh}
  % \state{Pennsylvania}
  \country{USA}
}

\author{Obigbemi Imoleayo Foyeke}
\email{iobigbemi@unilag.edu.ng}
\affiliation{%
  \institution{University of Lagos}
  \country{Nigeria}
}

\author{Assane Gueye}
\email{assaneg@andrew.cmu.edu}
\affiliation{%
  \institution{Carnegie Mellon University-Africa}
  \country{Rwanda}
}

\author{Edith Luhanga}
\email{eluhanga@andrew.cmu.edu}
\affiliation{%
  \institution{Carnegie Mellon University-Africa}
  \country{Rwanda}
}

\author{Alexander Rusero}
\email{ruseroa@africau.edu}
\affiliation{%
  \institution{Africa University}
  \country{Zimbabwe}
}

\author{Karen Sowon}
\email{ksowon@iu.edu}
\affiliation{%
  \institution{University of Indiana-Bloomington}
  \country{USA}
}

\author{Giulia Fanti}
\email{gfanti@andrew.cmu.edu}
\affiliation{%
  \institution{Carnegie Mellon University}
  % \city{Pittsburgh}
  % \state{Pennsylvania}
  \country{USA}
}

\renewcommand{\shortauthors}{Agadagba et al.}

\fi

%-------------------------------------------------------------------------------
\begin{abstract}
%-------------------------------------------------------------------------------
  \input{Sections/00-Abstract} 
\end{abstract}



 %% CCS: DO NOT REMOVE but you MAY update


\maketitle

\input{Sections/01-Introduction}

\input{Sections/02-Background}

\input{Sections/03-Methods}
\input{Sections/04-Results}

\input{Sections/05-Discussion}
\input{Sections/06-Conclusion}

\input{Sections/07-Acknowledgments}

%%-------------------------------------------------------------------------------
\ifdoubleblind
\else
\section*{Acknowledgments}
This work was supported in part by the Gates Foundation, 
the Initiative for CryptoCurrencies and Contracts (IC3), and National Science Foundation grant SaTC-2325477. The views
and opinions expressed in this study are those of the authors and
do not necessarily reflect the views or positions of the sponsors.
\fi
%%-------------------------------------------------------------------------------
%
%The USENIX latex style is old and very tired, which is why
%there's no \textbackslash{}acks command for you to use when
%acknowledging. Sorry.
%
%\textbf{Do not include any acknowledgements in your submission which may deanonymize you (e.g., because of specific affiliations or grants you acknowledge)}
%
%-------------------------------------------------------------------------------
% optional clearing of the page
% \cleardoublepage

% \ifdoubleblind
% \appendix
% % \input{Appendices/01-open-science}
% \input{Appendices/00-ethics}
% \fi

% \cleardoublepage
%\bibliographystyle{plainurl}
\ifdoubleblind
\newpage
\bibliographystyle{ACM-Reference-Format}
\else
\bibliographystyle{plainnat}
\fi
\bibliography{Bibliography/Bibliography}

\ifdoubleblind 
\appendix
\input{Appendices/00-ethics}

\else
\appendix

\input{Appendices/00-ethics}
\fi

\input{Appendices/02-interview}
\input{Appendices/03-codebook}

%%%%%%%%%%%%%%%%%%%%%%%%%%%%%%%%%%%%%%%%%%%%%%%%%%%%%%%%%%%%%%%%%%%%%%%%%%%%%%%%
\end{document}

%% file: Sections/00-Abstract.tex
Digital payment systems are a cornerstone of African consumer finance,
% , there exist many types of digital
% payment systems, including 
including money transfer applications, mobile money,
cryptocurrencies, stablecoins, and central bank digital currencies (CBDCs). 
While
many studies explore how people use digital
payment systems (in Africa and beyond), little is known about how people \emph{choose} and reason about 
categories of payment systems.
% , and how they view the
% tradeoffs between different categories. 
We conducted
qualitative interviews in three African countries---Nigeria, Tanzania, and Zimbabwe---to understand how and why people use various payment systems.
% , and what
% influenced them to start using these systems. 
Our study suggests that
users navigate accounts and payments on multiple payment systems through a complex selection process 
% using different platforms for different types of payments 
% several notable findings  regarding 
based on perceived tradeoffs between utility, privacy, and security. 
Many users trust government issuers to protect them from
scams, but they do not trust those same institutions to build reliable products or prioritize customer satisfaction.
% This selection process is driven in part by financial considerations, but also by security, privacy, and trust preferences.
% Users also want human agents for conflict
% resolution, but do not trust humans in centralized bodies to handle
% their information securely and privately, and thus default to solutions
% that are perceived as decentralized, such as cryptocurrencies and
% stablecoins. 
Our findings suggest compelling directions for
regulators and the research community to design systems that 
balance users' trust and utility needs.

%% file: Sections/01-Introduction.tex
% Introduction
\section{Introduction}

The proliferation of digital payment systems in Africa
% including mobile money, cryptocurrencies, and Central Bank Digital Currencies (CBDCs), 
is rapidly transforming financial ecosystems and contributing to growing financial inclusion across Africa \cite{soutter2019digital}. Mobile money platforms in particular, like M-Pesa and Ecocash, have bridged gaps in traditional banking, enabling millions in rural and urban areas to access digital financial services \cite{fabregasMobileMoneyEconomic2022}. 
The African digital payments market was predicted to 
% increase by 152\% from 2020 to 
reach \$40 billion by 2025 in domestic payments alone, with transactions growing to 188 billion in volume \cite{McKinsey2025}. 
Such access to financial services is foundational for poverty reduction and economic growth in Sub-Saharan Africa \cite{Osuma2025}. 

% Indeed,  adoption of digital payment systems in Africa has steadily gained momentum since the year 2000, with a record increase around 2020 due to the COVID-19 pandemic \cite{HeresWhyAfrica2023a}. 
In addition to  ``traditional'' digital payment systems (e.g., money transfer applications,  mobile money), African users and innovators have embraced  several emerging technologies, including central bank digital currencies (CBDCs), cryptocurrencies, and stablecoins \cite{McKinsey2025}. Nigeria, for instance, introduced the eNaira CBDC in 2021 as a pioneering effort to improve transaction efficiency and financial oversight of its monetary system \cite{omotuboraSameNairaMore2024}. Other countries, such as Zimbabwe, have also deployed CBDCs \cite{UAEsDigitalDirham2025}, with more actively exploring pilot releases \cite{bog-cbdc}. Kenya, Nigeria and South Africa, meanwhile, are ranked among the top users of cryptocurrencies globally \cite{domorWhySubSaharanAfrica2025}. 

Although Africa's adoption of digital payment systems is anticipated to increase as more people become connected to digital infrastructure \cite{ImpactDigitalLiteracy}, \emph{adoption is likely to be unevenly distributed} across the continent and across  payment methods. This imbalance could stem from factors including cost, user trust in service providers, ease of use, technological access, regulatory frameworks, and social influence, all of which play critical roles in shaping adoption. For example, while mobile money thrives in most parts of Africa \cite{vanteutem}, CBDC adoption faces hurdles due to low public awareness and privacy and security concerns \cite{prodanRisePopularityCentral2024}. \textbf{Understanding reasons for uneven adoption and use of payment systems is essential for the rollout and deployment of new systems.}

Today, there exist many primary-source studies---both in Africa and globally---on user adoption and attitudes toward \emph{individual categories} of digital payment systems. 
% For example, Ricci \emph{et al.} investigated CBDC issuance and adoption and challenges in Africa \cite{ricciCentralBankDigital2024a},   while Mazambani and Mutambara quantitatively studied cryptocurrency adoption in South Africa \cite{mazambani2020predicting} (more examples in \Cref{sec:related}).
% \gf{It would be  better if we phrased as "
For example, researchers have evaluated attitudes and usage of CBDCs   \cite{bosua2024public,ricciCentralBankDigital2024a,bijlsma2021triggers,ngo2023governance}, cryptocurrencies \cite{zhang2019security,do4africaCryptocurrencyAdoption,mutiso2020assessment,mazambani2020predicting}, and mobile money (MoMo) \cite{mothobi2017infrastructure,akinyemi2020determinants,sowon2024role,luhanga2023user,kotut2024things}. 
% Can we get something like this?}
However, to our knowledge, there have not been studies \emph{comparatively} examining user perspectives between various digital payment methods.
% across multiple African countries. 
Moreover, many of the existing studies are structured  surveys \cite{akinyemi2020determinants,mothobi2017infrastructure,ngo2023governance,mutiso2020assessment,bijlsma2021triggers,mazambani2020predicting}, which may inhibit a deeper understanding of peoples'  attitudes and reasoning.

We address this gap in the literature through qualitative interviews with users of five categories of digital payment systems---banking applications, MoMo, cryptocurrencies, stablecoins, and CBDCs---in three Sub-Saharan African countries---Nigeria, Tanzania, and Zimbabwe.  
% The significance of our work lies in understanding the dynamics that are critical for addressing challenges that hinder equitable adoption of digital payment systems in Africa thus closing this gap in literature. 
Our study investigates 
% users in Tanzania, Nigeria, and Zimbabwe, outlining the 
enablers and barriers to digital payment adoption and continued use. 
% , exploring user experiences, demographic influences, and decision-making processes toward the adoption and continued use of particular digital payment methods. 
In particular, we ask the following research questions:

\begin{enumerate}[label=RQ\arabic*]
    \item What factors affect users' \textbf{initial adoption} of new digital payment systems?
    \item What factors (e.g., trust in platforms/supporters, utility, ease of use, recourse mechanisms, and popularity/understanding) significantly influence users’ \textbf{continued use} of digital payment systems?
    \item What \textbf{failures, risks, and breakdowns} do current users face after adoption? %and how do these differ across digital payment system categories? 
\end{enumerate}
For each research question, we explore how answers differ across digital payment system categories.
Our interviews highlight several important findings. 
\begin{enumerate}
    \item While participants were drawn to payment systems that display reliability, efficiency, and convenience, adoption was not always entirely voluntary: institutional requirements, cash scarcity, monetary instability, and other economic sometimes made adoption a practical necessity rather than a choice. Most of our participants also reported significant usability challenges with these payment systems, including infrastructure failures and failures of human systems (e.g., lack of pathways for recourse).
    \item Participants displayed nuanced, sometimes conflicting attitudes about trust in government bodies.  These attitudes impacted users' willingness to adopt and use government-issued payment systems.
    \item Users engage in a complex decision-making process to decide which payment system to use for which transactions, and dynamically reallocate financial activity across systems.
    These decisions can vary with timing and transaction details, and they are 
    enabled by a rich environment of digital payment platforms, each with their own perceived shortcomings. 
    % Circumstances often led users to , rather than simply choosing among systems for individual transactions. 
\end{enumerate}

We conclude with recommendations 
% By providing an analysis of sentiment across these three countries, the research seeks to offer actionable insights 
for policymakers, financial institutions, and researchers. 
% to advance equitable financial inclusion throughout Africa.
In particular, we recommend that regulators encourage interoperability for the increasingly fragmented African digital payment ecosystem, 
% \gf{Is this consistent with the new recommendations?}
% These factors include organic social influence, paid advertisements, and financial incentives, but they also include careful release timing and designing a product in response to public sentiment around \emph{existing} offerings. 
and that the HCI research community further study the impacts of design recommendations for digitally illiterate users, including delegated transactions. 
% Our study touches on the growing technological sophistication required to navigate digital payment systems; this suggests a commensurate growth in the divide between digitally literate and illiterate users.

% Such studies can inform future governmental efforts, particularly how to prioritize resources. 

% Overall, our study provides a comprehensive window into how users of various digital payment systems think about their relative benefits and downsides in three African countries. 
% These results could help inform the digital payments industry (particularly in Africa) understand how to better meet the needs of their users.  

% The remainder of the paper is organized into six sections. Section II examines the background and related work. Section III gives a concise overview of the methodology, focusing on data collection, analysis, and ethical considerations of this study. Section IV presents results and qualitative discussion of the findings, while Section V concludes the paper.

% \subsection*{Research Questions}

%% file: Sections/02-Background.tex
\section{Background and Related Work}
\label{sec:related}
We first provide background context on the five categories of digital payment systems that feature prominently in our study: mobile money, money-transfer applications, cryptocurrencies, stablecoins, and central bank digital currencies (CBDCs). 
\Cref{tab:examples} lists examples from each category of payment system in our three countries of interest.
% For each category, we first describe how the system works, focusing on key actors and transaction flows, review its adoption in Africa, and then review prior related research drawing on both global and Africa-focused studies. While a substantial body of work examines these payment systems in isolation, comparatively little research considers how users navigate and evaluate multiple digital payment technologies in parallel.

\begin{table*}[t]
\centering
\begin{tabular}{cccccc}\toprule
                  & \textbf{\begin{tabular}[c]{@{}c@{}}Mobile\\ money\end{tabular}} & \textbf{Cryptocurrency} & \textbf{Stablecoins} & \textbf{\begin{tabular}[c]{@{}c@{}}Money transfer\\ applications\end{tabular}} & \textbf{CBDCs}                                                \\ \midrule
\textbf{Nigeria}  & \begin{tabular}[c]{@{}c@{}}OPay, Palmpay, \\ Kuda \end{tabular}                                                             & \begin{tabular}[c]{@{}c@{}}Bitcoin, Litecoin \\ Ethereum\end{tabular}       & USDT, USDC           & \begin{tabular}[c]{@{}c@{}}Western Union, Remitly \\ Flutterwave \end{tabular}                                                                   & eNaira                                                        \\ 
\hline
\textbf{Tanzania} & \begin{tabular}[c]{@{}c@{}}M-Pesa, \\ Tigo Pesa \end{tabular}    & \begin{tabular}[c]{@{}c@{}}Bitcoin, Litecoin \\ Ethereum\end{tabular}  & USDT, USDC           & \begin{tabular}[c]{@{}c@{}}Western Union \\ MoneyGram\end{tabular}                                                                  & ---                                                           \\ 
\hline
\textbf{Zimbabwe} & EcoCash                                                       & \begin{tabular}[c]{@{}c@{}}Bitcoin, Litecoin \\ Ethereum\end{tabular}        & USDT, USDC           & \begin{tabular}[c]{@{}c@{}}Western Union, InnBucks \\ MoneyGram\end{tabular}                                                                  & \begin{tabular}[c]{@{}c@{}}Zimbabwe\\ Gold (ZiG)\end{tabular} \\ \bottomrule
\end{tabular}
\caption{Examples of popular products or offerings in the five categories of digital payment systems we study in this work.}
\label{tab:examples}
\end{table*}

\subsection{Mobile money (MoMo)}
MoMo is an account-based digital payment service operated primarily by mobile network operators and used by both 
% banked and unbanked users, including both 
smartphone and basic-phone users. 
% Unlike mobile banking systems, 
MoMo does not require a formal bank account, requiring only a user’s phone number as an identifier; transactions are conducted through USSD menus, SIM-based interfaces, or mobile applications.
%\subsubsection{How Mobile Money Works}
%Common MoMo transactions include deposits (cash-in), withdrawals (cash-out), person-to-person (P2P) transfers, and person-to-merchant payments.
MoMo services  allows users to convert cash to/from electronic value (cash-in and cash-out, respectively) and P2P digital value transfer among participants.
% , and convert electronic value back into cash (cash-out). 
% During cash-in, physical cash is exchanged for e-money credited to a user’s mobile wallet. 
% during cash-out, e-money is debited from the wallet and exchanged for physical cash. 
% Once value is held digitally, users can perform person-to-person (P2P) transfers by sending funds to another  phone number.
% after which recipients may retain the balance digitally or withdraw cash. 
MoMo operations rely on dense networks of transactional agents, typically small local businesses \cite{sowon2024role}, 
% Human agents maintain e-wallets with higher balance limits than user wallets, enabling the exchange of physical cash and e-money.
% between the user and the agent. Through 
% This model enables financial transactions 
allowing systems and users to circumvent limited banking infrastructure. 
% while remaining accessible to users with basic mobile devices.

% \paragraph{Adoption}
MoMo has significantly improved financial inclusion in Africa due to its simple design and popularity \cite{avom2023financial}.
Today, Sub-Saharan Africa has an estimated 165 live mobile money services, the majority operated by mobile network operators (MNOs), serving  over 1.1 billion registered accounts \cite{momo2025}.

%\subsubsection{Adoption and Use in Africa}
\paragraph{Related work}
% Two strands of literature on Mobile Money relate to our work. The first studies adoption of MoMo. 
% Several studies have explored enablers and barriers influencing adoption across Africa. %showing that user demographics [CITE], technological factors (e.g., ease of use) [CITE] and contextual factors (e.g., social influence [CITE] influence adoption and use.
Several studies have explored the applicability and adoption of MoMo, both in African and globally \cite{hinman2010opportunities,akinyemi2020determinants,mustafa2019digital,kotut2024things}. 
% Akinyemi \emph{et al.} investigate the determinants of MoMo adoption in rural Africa using data from ten countries \cite{akinyemi2020determinants}. 
Akinyemi \emph{et al.} find that %demographic factors such as 
age, education, income, employment status, and mobile phone ownership are key drivers of both adoption and usage with younger, more educated and high income individuals who own mobile phones being significantly likely to adopt and continue using MoMo. Similar studies have shown that perceptions of ease of use and usefulness shape mobile money adoption
% , with trust and perceived risk also influencing adoption in some contexts 
\cite{mndeme2025enablers,ndekwa2018adoption}, 
% In Akinyemi et al \cite{akinyemi2020determinants} %The study also highlights the importance of user perception showing that favorable view of MoMo positively influence on adoption: A 
% a large majority of respondents perceived MoMo as faster (89\%), easier (89\%), safer (83\%), and more convenient (80\%) than alternative payment methods. %Other similar studies have shown that factors  including perceivedSubsequent studies similarly emphasize perceived value, affordability, and compatibility with everyday financial practices as important for sustained use, alongside 
% Contextual factors also play a role in adoption; 
and users tend to adopt MoMo when other trusted people in their social spheres use it \cite{bitrus2021mobilepayment,ndekwa2018adoption}. Barriers to adoption include high transaction fees and challenges related to low merchant acceptance of cashless payments in some contexts %[CONFIRM that citations talk about barriers]
\cite{lissah2024cashless}. %\tk{DONE - Lissah et al strongly support this claim as it talk about factors that affect the adoption of cashless payment is highly influenced by perceived price value of the service...}.
%\ks{Edited the paragraph from what was here. Add a few more relevant studies on adoption before moving to those on Privacy/Security}

%Several studies have been done focusing on digital payment systems. Our related work is divided into four sections discussing prior work on (1) Mobile Money, (2) Cryptocurrencies, (3) CBDCs, and (4) Money transfer applications.

%\textbf{Privacy, Security, and Trust.}
Another strand of related literature concerns %Mobile money research in Africa consistently identifies 
usability, privacy, and security of MoMo \cite{harris2013, makulilo2015privacy,sowon2024role,luhanga2023user,sowon2025}. Sowon \emph{et al.} \cite{sowon2024role} highlighted usability concerns and challenges related to KYC, revealing several workarounds that users adopt facilitated by their choice of agent, e.g., preferring agents with lax KYC practices.
% who allowed transactions without requiring users to present any identification as required by the system. 
In a related study, Luhanga \emph{et al.} also revealed concerns around information sharing with agents, leading to   workarounds such as registering MoMo accounts under another person's ID \cite{luhanga2023user}.

\subsection{Money transfer applications}
Money transfer apps typically digitize existing financial infrastructure—most often banks and other fintechs—rather than replacing them entirely like MoMo. Fintech start-ups in Africa such as Flutterwave and Chipper Cash are built atop  existing existing financial and banking systems to enable mobile payments and transfers \cite{techdeskAfricasFintech}.
% , extending digital financial access through bank integrations and APIs \cite{techdeskAfricasFintech}. 
% remittances by coordinating transfers across existing banking infrastructures . 
% In Africa, these services are typically operated by regulated financial institutions or fintech companies \cite{cafonBankingOperations}. 
% Applications  are commonly used for person-to-person transfers, 
% allowing users  to send and receive money instantly, usually via smartphones and the internet \cite{cafonBankingOperations}. 
Using an app linked to their bank account, 
% a user initiates a transaction by entering the recipient's account number or phone number and the amount to transfer. The money is 
users can instantly send payments through the banking system 
% with both sender and recipient receiving confirmation. 
% The recipient can 
% transact in the app or 
or withdraw cash through an agent or an ATM, as in MoMo. 
% While the user experience feels like wallet-to-wallet transfers, transactions are actually bank-to-bank in the background, with 
% Agent networks and POS terminals enable cashing-in and -out, similar to MoMo.
% thus extending financial inclusion without the reliance of bank branches---similar to MoMo. 
% This model differs  from classic MoMo systems (e.g., M-Pesa), which are led by telecom operators, tied directly to SIM cards, and designed to work without bank accounts, smartphones, or internet access. 
Unlike classic MoMo systems, money-transfer applications often rely on users' bank accounts and cards or mobile money wallets  \cite{cafonBankingOperations}. 
% While users favor fintech platforms for their convenience and accessibility, fraud and trust concerns remain prevalent \cite{isiaku2024mobile}. 

%\subsubsection{Adoption and Use in Africa}
%Across Africa, money transfer applications have expanded alongside the growth of fintech ecosystems and real-time payment infrastructure, supporting domestic payments and remittance flows. Industry and policy reports document rapid uptake driven by speed and convenience, particularly in countries with active fintech sectors \cite{techdeskAfricasFintech}.

% \paragraph{Adoption}
{Nigeria's} digital payment ecosystem features prominent use of money transfer platforms \cite{eib2024fintech}. Nigeria reported 7.9 billion dollars worth of transactions in 2024 \cite{Egobiambu2025DigitalPaymentsNigeria} on platforms such as Paystack and Flutterwave. 
%Channels Television (2025) documented that Nigeria processed 7.9 billion real-time transactions in 2024, more than any other African country, driven by instant payment infrastructure, including the Nigeria Inter-Bank Settlement System (NIBSS) and widespread POS terminal deployment. 
%like OPay, PalmPay, and MTN MoMo
%perceived risk and security concerns continue to shape trust and usage intentions, highlighting that fintech adoption in Nigeria is driven by convenience and accessibility but constrained by unresolved fraud and trust anxieties. 
Money transfer applications like Western Union and MoneyGram are also popular in other African countries, including Zimbabwe and Tanzania \cite{munthali2024impact}.
% In other countries, available apps depend on fintech offers and factors such as user base, popularity, availability of POS merchants and transaction costs 
% [CONFIRM is this is true]
%In Nigeria, platforms such as OPay and Flutterwave are widely used through integration with bank and merchant payment infrastructure \cite{cafonBankingOperations}. These applications are commonly used for person-to-person transfers and merchant payments, extending digital financial access through mobile-first interfaces built on top of existing financial systems rather than replacing them entirely \cite{techdeskAfricasFintech,cafonBankingOperations}.

\paragraph{Related work}
%\ks{This section is conflating MoMo and Money Transfer Apps. I commented out some sections. We need literature on Money Transfer Apps}

%\ks{Studies on money transfer applications have focused on their impact on development through increased remittances}
Several studies show how users engage with money transfer applications around the world \cite{lissah2024cashless, hillman2014user, olaleye2017users}. Hillman \emph{et al.} note that vendor-specific payment applications offer users benefits such as faster transaction completion and suitability for everyday purchases, as well as the ability to circumvent certain regulatory constraints \cite{hillman2014user}. At the same time, users express persistent concerns around privacy, a theme that has continued to surface in later studies of peer-to-peer payment platforms like Venmo \cite{tandon2022know}. The dynamics of adoption and use differ substantially. Studies in China and India document settings in which cashless payments are increasingly required by merchants or expected by customers, hence driving adoption of money transfer applications \cite{shen2020can, jain2021protecting}. In Nigeria, anxieties around application use and concerns about privacy negatively shape experiences with digital payment systems \cite{olaleye2017users}.

\subsection{Cryptocurrencies and stablecoins}
Cryptocurrencies are digital assets recorded on distributed ledgers that enable peer-to-peer value transfer without reliance on centralized intermediaries such as banks. Most cryptocurrencies rely on a blockchain, a decentralized ledger that orders and validates transactions %across a network of participants 
\cite{zhang2019security}. 
% Transactions are created and signed using cryptographic keys, broadcast to the network, and included in blocks that become part of an immutable chain once consensus is reached; after confirmation, transactions are generally irreversible \cite{zhang2019security}. 
While cryptocurrencies are globally accessible, they require users to have digital wallets connecting cryptocurrencies to local fiat currencies. Converting between cryptocurrencies and local fiat currencies typically requires centralized exchanges or informal brokers. 
% which serve as on-ramps (fiat-to-crypto) and off-ramps (crypto-to-fiat).

Stablecoins are a class of cryptocurrencies designed to maintain a stable value relative to a %reference asset, most commonly a 
fiat currency such as the U.S. dollar. They are often used as a medium of exchange or store of value %in contexts where 
when local currencies are volatile or when users seek fast settlement.
% than traditional banking systems allow. 
While many stablecoins operate on blockchains like cryptocurrencies, 
%they differ from other cryptocurrencies in 
that their value stability depends on issuer mechanisms or collateral arrangements \cite{Garita2024Stablecoins}.
Due to these differences, we treat stablecoins as a separate category in our study.

% \paragraph{Adoption}
Through mid-2025, Sub-Saharan Africa recorded approximately \$205 billion in on-chain cryptocurrency transactions \cite{dmarketforcesAfricas205B}. Traditional cryptocurrency adoption is particularly strong in Nigeria, Kenya, South Africa, and Zimbabwe \cite{do4africaCryptocurrencyAdoption}, where many people trade and hold crypto assets via mobile devices and peer-to-peer platforms despite regulatory uncertainty and mixed government stances on digital asset legality. Nigeria, Kenya, and South Africa also rank among the top 10 countries globally for cryptocurrency adoption \cite{bitcoinUkraineRussia}. 
% with Bitcoin comprising 89\% of crypto purchases in Nigeria and 74\% in South Africa \cite{chainalysis2025subsaharan}.
% (significantly higher than the 51\% share in USD markets.
% Among Nigerian crypto investors in 2022, Bitcoin remained dominant at 76\%, followed by Ethereum at 50\% and Binance Coin at 45\%.
% In response to regulatory restrictions, 
% adoption has remained high, with users sometimes moving 
% some users have moved to peer-to-peer trading, which complicates oversight \cite{oladipupo2023effects,wallchartafrica2024cryptocurrency}.

Stablecoins represent a significant fraction of African cryptocurrency holdings: in Nigeria they make up 43\% of the region's crypto transaction volume, up from roughly 30\% in 2023 \cite{dmarketforcesAfricas205B}, driven primarily by remittances and inflation hedging \cite{coingeek2024stablecoins}. 
Sending a \$200 remittance from Sub-Saharan Africa was reported to cost $\sim 60$\% less when using stablecoins compared to traditional remittance methods \cite{chainalysis2024}.

\paragraph{Related Work} 

%\textbf{Security, Privacy, and Trust} 
% Prior work on cryptocurrency adoption 
% % finds that demographic factors and user perceptions play a significant role in uptake. Studies 
% consistently shows that cryptocurrency users and investors tend to be younger and predominantly male \cite{jalan2023trust}. Urban residence and higher levels of education, particularly tertiary education, are also associated with higher rates of cryptocurrency adoption \cite{tgm2024nigeria}. 
% Beyond demographics, perceptions of trust and distrust toward centralized institutions—including banks and governments—are key drivers of cryptocurrency use \cite{shahzad2024cryptocurrency} \cite{kumar2024drivers} \cite{sharma2023empirical} \cite{adaramola2025dark}. While decentralization and limited government regulation are often cited as motivations for adoption, other studies suggest that these properties are not uniformly valued. 
% Adaramola \cite{adaramola2025dark} finds that users may still prefer cryptocurrency platforms that offer recourse mechanisms or some degree of regulatory oversight, highlighting tensions between ideals of decentralization and practical risk tolerance. 
Prior work on cryptocurrency adoption consistently shows that users and investors tend to be younger, predominantly male \cite{jalan2023trust}, urban, and more educated, particularly with tertiary education \cite{tgm2024nigeria}. Beyond demographics, perceptions of trust and distrust toward centralized institutions—including banks and governments—are key drivers of cryptocurrency use \cite{shahzad2024cryptocurrency,kumar2024drivers,sharma2023empirical,adaramola2025dark}. While decentralization and limited government regulation are often cited as motivations for adoption, other studies suggest these properties are not uniformly valued: Adaramola \cite{adaramola2025dark} finds that users may still prefer platforms offering recourse mechanisms or some degree of regulatory oversight. 

Stablecoins demonstrated partial, context-dependent hedging effectiveness across high-inflation economies, particularly where weak monetary institutions and currency devaluation prevail. While they offer high mobility and decentralized access, they suffer from regulatory uncertainty. In Nigeria, Kenya, and Venezuela, stablecoins enabled citizens to bypass capital controls and acquire savings more easily than through traditional banking \cite{akomolehin2025cryptocurrencies}. %The study concluded that stablecoins should be treated as complementary rather than substitute hedging tools within diversified investment strategies.

\subsection{Central Bank Digital Currencies}
Central Bank Digital Currencies (CBDCs) are digital currencies issued and backed by central banks, 
% They are implemented using centralized or hybrid ledger architectures in which the 
% The central bank issues digital currency 
either directly to users or through regulated intermediaries such as banks or payment service providers. 
% Like other digital payment systems, 
Users access CBDCs via digital wallets that allow them to send and receive value electronically. 

CBDC adoption has  been low in Africa, with Zimbabwe and Nigeria being the only 2 countries that have deployed CBDCs (ZiG and eNaira, respectively); however, African countries like Rwanda are piloting or testing early-stage CBDCs \cite{RwandaPublishesNew2024}. 
Within Nigeria, eNaira adoption was reported to reach only $0.8\%\ $ of bank accounts by mid-2023 \cite{wezel2023nigeria}.
Zimbabwe saw higher adoption levels for ZiG, likely due to the government making ZiG the official national currency since 2024; nonetheless, ZiG remains unpopular in large swaths of the economy \cite{cornell2025zig}.

% Nigeria launched the eNaira in 2021, becoming the first African country to introduce a retail CBDC.  

%\subsubsection{Adoption and Use in Africa}

\paragraph{Related work}
Prior work has explored reasons for low CBDC adoption \cite{alberolaCentralBankDigitalb,alberolaCentralBankDigitalc,ricciCentralBankDigital2024a,oziliSurveyCentralBanka}, including a limited value proposition beyond existing payment systems, insufficient use cases (only 10\% of users completed 5-10 transactions), low merchant adoption, and concerns about bank disintermediation \cite{davidwest2023cbdc}. Security, privacy, and usability concerns also commonly reduced trust in the system \cite{oziliSurveyCentralBanka}, with the centralization of CBDCs heightening fears of surveillance \cite{kaur2024cbdc}.

% Trust in government also emerged as a significant barrier, with some users expressing concerns about governmental oversight into financial transactions. 
% Under current design, the central bank maintains visibility into all transactions, creating privacy concerns given broader distrust of government institutions. The Cornell Business School analyzed eNaira's struggles, identifying factors including mixed messaging after the cryptocurrency ban, implementation on private blockchain contradicting decentralization ethos, and tiered KYC requirements creating barriers for users lacking identification documentation \cite{cornell2023enaira}.

% \textbf{Infrastructure and Awareness.} Nigeria's electricity crisis and limited internet connectivity, especially in rural communities, poses as fundamental adoption barriers \cite{adebayo2024addressing}. The CBN has begun addressing offline access through USSD codes enabling transactions without internet-enabled phones. Many potential users do not understand how CBDC  differs from other money forms, and whether it will replace cash \cite{davidwest2023cbdc}.

% \textbf{Comparative Studies.} 
% Several recent empirical studies have applied partial least squares structural equation modeling (PLS-SEM) and machine-learning techniques to survey data (e.g., ~600 respondents) on China’s e-CNY adoption, finding that technology-related perceptions such as 
Studies from China and India
% e-CNY CBDC adoption showed that
show that
perceived ease of use, attitude, security, and usefulness positively influenced intent to adopt \cite{liu2025cbdc,kaur2024cbdc}. 
% A randomized controlled trial found that 
Consumers shown short videos about CBDC features were substantially more likely to update beliefs and demonstrate adoption likelihood, though most declined options to learn more, suggesting passive awareness does not translate to active engagement \cite{ecb2024consumer}.
% Among Gen-Z  participants in India, behavioral intention to adopt CBDC was positively influenced by attitude, 
% % with perceived trust and security serving as mediating factors in key relationships, although 
% while effort expectancy had no significant effect on attitude \cite{kaur2024cbdc}.
% demonstrated a non-significant 

\subsection{Our contributions} While existing studies have examined individual payment technologies in isolation---e.g., cryptocurrency adoption drivers \cite{shahzad2024cryptocurrency, nguyen2025cryptocurrency}, mobile money expansion \cite{sowon2024role, sowon2025}, and CBDC implementation challenges \cite{davidwest2017agent}---few have conducted a comparative analysis across different categories of payment types within a single population. 
In the closest work to ours, Kotut and Alikhan \cite{kotut2024things} run qualitative interviews in Kenya exploring users' ownership of multiple devices (including SIM cards) and mobile money accounts.
Our observations parallel some of theirs, including that people use different accounts and devices to separate financial transactions. 
Relative to \cite{kotut2024things}, our study covers more/different countries (Nigeria, Tanzania, Zimbabwe)  and explicitly examines relations between different payment \emph{categories} 
% In particular, \cite{kotut2024things} does not mention 
including cryptocurrency-based payment systems and CBDCs. 
Our findings also surface complex dynamics around trust in government and service providers that are not explored in \cite{kotut2024things}.

% Existing quantitative research on digital payment systems relies on survey instruments measuring stated intentions rather than actual usage behaviors. This study addresses these gaps through qualitative interviews with active users,
% % across Nigeria, Tanzania, and Zimbabwe, 
% examining how contextual factors like trust, infrastructure, and regulatory environment differentially shape adoption across payment technologies. 
% We contribute a theoretically grounded analysis of how information flow norms influence user comfort with  digital payment systems.

%% file: Sections/03-Methods.tex
\section{Methods}

This study employed a qualitative research approach to investigate digital payment adoption and perceptions 
% The research utilized a two-phase design, beginning with preliminary data collection through surveys focused on cryptocurrency adoption and perceptions, which subsequently informed the development of a more comprehensive interview-based study examining broader digital payment methods.
 % We conducted interviews 
 in three Sub-Saharan African countries: Nigeria, Tanzania, and Zimbabwe. All three countries are fairly advanced in terms of financial technology adoption, with widespread MoMo and cryptocurrency adoption \cite{chainalysis2025subsaharan,mothobi2017infrastructure,opay,PalmPay,Egobiambu2025DigitalPaymentsNigeria,mutiso2020assessment,abdurrahaman2024revisiting}. Two of the countries---Nigeria, with eNaira, and Zimbabwe, with Zimbabwe Gold (ZiG)---have an active CBDC, and one (Tanzania) does not. 
 % we wanted to understand if the absence of a deployed CBDC would change results. 
 % based on established researcher connections within these regions and each country's particularly dynamic digital payment landscape. For example, Nigeria’s recent introduction of the eNaira in 2021 and rapid adoption of mobile money systems post-COVID. [Include examples of each country's digital payment landscape here.] 
 These countries provide varied economic and technological contexts for examining regional digital payment adoption patterns.

 % \gf{Nicolas, is it ok to just point to the appendix or do we need a section in the main paper?}

\subsection{Study design and pilots}

Our interview script, data collection procedures, and recruitment
and screening material were developed primarily by three lead researchers; two are natives of Tanzania and Zimbabwe, respectively, and one is a first-generation immigrant from Nigeria. 
Feedback was provided by a team of co-authors with complementary expertise, including advisors from each of the three countries who guided the ethics approval, recruitment, and local arrangements.
% We used semi-structured interviews to maintain
% consistency across three countries. 

The primary data source for this study consisted of in-depth, semi-structured interviews focusing on five main categories of digital payment methods: mobile money systems, stablecoins, central bank digital currencies (CBDCs), money transfer platforms, and traditional digital payment methods. Before beginning the main  interview, each participant was requested to fill out a demographics sheet that covered what particular digital payment services they use or have used previously, as well as (optionally) their demographic information, including their country of origin,  country of residence, age bracket, education level, gender, occupation, and rural or urban residence. Once completed, participants were asked interview questions regarding their digital currency usage for each service they selected; users were asked to compare the aforementioned services, and they were asked to react to hypothetical scenarios involving the adoption of new digital payment systems (interview in Appendix \ref{app:interview}).
% Full interview questions are included in Appendix \ref{app:interview}.

Prior to full data collection, seven pilot studies were conducted from March 2025 to July 2025 to refine the interview protocol, 
% These pilot interviews were used to 
optimize question flow, ensure full understanding of language and concepts, and shape interview structure and duration. 
The iterative process of pilot testing allowed for refinement of the interview script to maximize response quality and participant engagement. 
The results from these  pilot interviews are not
presented here.

The lead researcher from Tanzania  translated the interview guide to
Swahili, the official language of Tanzania; the translation was certified by a local professional prior to obtaining local ethics approval.
All interviews in Tanzania were conducted in Swahili for participant comfort and participation.
The interviews in Nigeria and Zimbabwe were conducted in English, which is an official language of both countries.

\begin{figure}[t]
    \includegraphics[width=0.9\columnwidth]{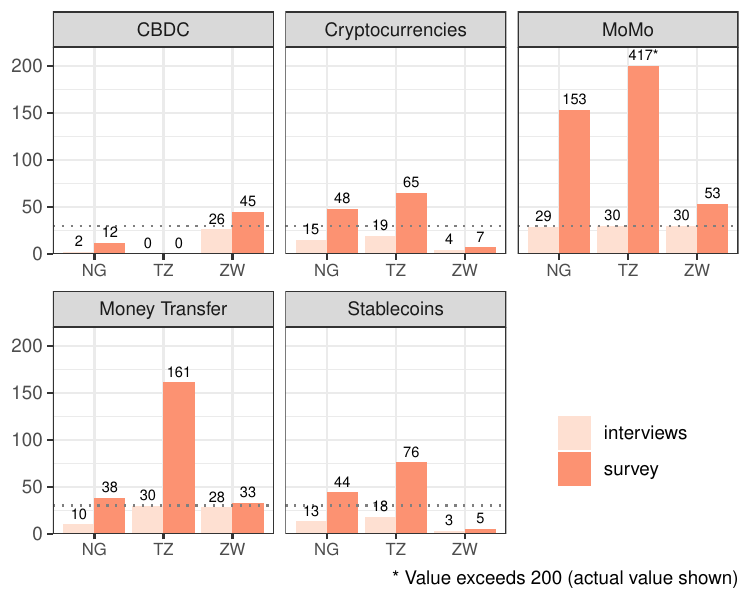}
    \caption{\label{fig:instruments} \textbf{Payment instrument usage} from our intake survey and  final interviews. For Tanzania, we limited the vertical bar to 200 for graph readability. The dotted line reflects the total number of participants in our interviews per country.}
\end{figure}

% \begin{figure}[h]
% %    \begin{tikzpicture}
% %        % Include your image
% %        \node[anchor=south west,inner sep=0] (image) at (0,0) {
% %
%         \includegraphics[width=0.95\columnwidth]{Figures/gender_balance}
% %        };
% %        % Draw red box overlay
% %        % Adjust coordinates (x1,y1) and (x2,y2) to position the box
% %        \draw[red, line width=2pt] (4,1.9) rectangle (5.75,3.2);
% %        \draw[red, line width=2pt] (4,5.9) rectangle (5.75,7.1);
% %    \end{tikzpicture}
% %
% % it's now annotated in the plot itself for better robustness
%     \caption{\label{fig:gender} \textbf{Gender balance} across payment instruments from our intake survey (right) and final interviews (left). For cryptocurrencies and stablecoins in particular, we had more male respondents on the intake survey than female (up to 3$\times$ as many, e.g., see red rectangles). }
% \end{figure}

\subsection{Participant recruitment and sampling}

90 participants were recruited across the three countries, with 30 participants per country. 
We gathered initial interest through an online intake form, advertised through physical and digital channels (e.g., WhatsApp groups, flyers) in the cities where the interviews took place, with assistance from our local liaisons.
It recorded each respondent's contact details along with their age, gender, education, residence, and payment categories used. In total, 623 respondents completed it (153 Nigeria, 417 Tanzania, 53 Zimbabwe). Beyond recruitment, the intake form served as a secondary source of descriptive data on the broader user pool. We report these counts alongside interview counts in Figures 1–3. The survey is a non-representative convenience sample and does not underpin our qualitative findings.

\begin{figure}[t]
    \includegraphics[width=0.9\columnwidth]{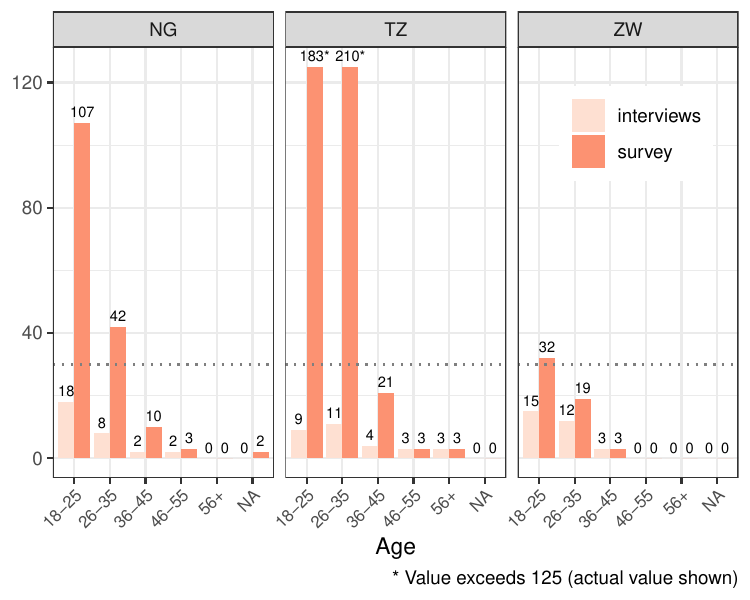}
    \caption{\label{fig:age} \textbf{Age breakdown} from our intake survey and  final interviews.  For Tanzania, we limited the vertical bar to 125 to preserve graph readability. The dotted line reflects the total number of participants in our interviews per country. Across all three countries, our respondents skewed heavily towards younger demographics. This could be due to a combination of our snowball sampling methods and age distributions that skew young in  the countries studied \cite{age-worldbank}.}
\end{figure}

\begin{figure}[h]
%    \begin{tikzpicture}
%        % Include your image
%        \node[anchor=south west,inner sep=0] (image) at (0,0) {
%
        \includegraphics[width=0.95\columnwidth]{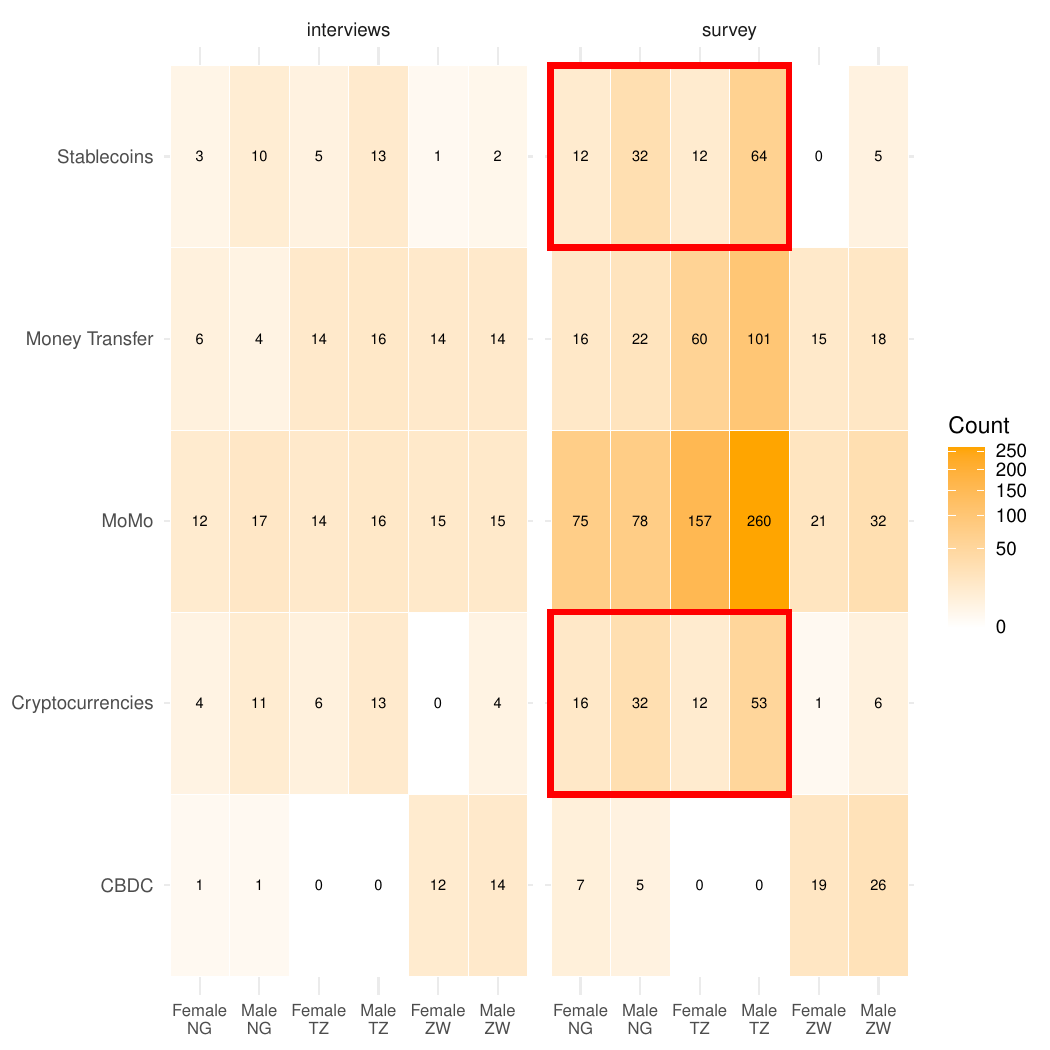}
%        };
%        % Draw red box overlay
%        % Adjust coordinates (x1,y1) and (x2,y2) to position the box
%        \draw[red, line width=2pt] (4,1.9) rectangle (5.75,3.2);
%        \draw[red, line width=2pt] (4,5.9) rectangle (5.75,7.1);
%    \end{tikzpicture}
%
% it's now annotated in the plot itself for better robustness
    \caption{\label{fig:gender} \textbf{Gender balance} across payment instruments from our intake survey (right) and final interviews (left). For cryptocurrencies and stablecoins in particular, we had more male respondents on the intake survey than female (up to 3$\times$ as many, e.g., see red rectangles). }
\end{figure}

From these intake respondents, purposive sampling was used to ensure demographic diversity for gender,\footnote{As nonbinary genders are not officially recognized in any of the three countries we studied 
\cite{nglgbtq,tzlgbtq,zimlgbtq}, 
our intake survey collected only binary gender options (these surveys had to be approved by the local ethics boards, some of which have government affiliations). 
However, we accounted for individuals who identify with other genders by making our gender designation 
question at the time of the interview optional.}
% } 
age range, education level, and use of digital payment systems. 
As shown in \Cref{fig:age,fig:instruments,fig:gender}, some demographics had much higher representation than others. 
Notably, we found relatively few CBDC users, compared to MoMo, which most respondents had used in some form (\Cref{fig:instruments}).
Male cryptocurrency/stablecoin users also outnumbered female users on our intake survey (\Cref{fig:gender}).
Finally, most of our respondents were 35 or younger (\Cref{fig:age}). 
To vary the demographics of our participants, we sampled the respondents to balance some of these attributes. We used most or all of the CBDC users on our intake survey in Zimbabwe and Nigeria, respectively, and we selected older respondents on the intake form, though some were later disqualified as they had misunderstood the requirements of the survey. 
 % of participants was applied to short intake questionnaires that prospective participants completed to express interest in the study. Interviewees were selected based on digital payment usage, availability for in-person interviews, and the target demographics we aimed to reach. This was done to ensure a diverse sample of participants. 

\subsection{Interviews}
\label{sec:interviews}
Individual interviews were conducted with each participant, lasting between 45 minutes and an hour and a half depending on the number of digital payment system categories they had experience with. All participants received monetary compensation of (the local equivalent of) 10 USD, paid via MoMo or cash when MoMo was inaccessible or inconvenient, plus up to 10 USD for travel costs to the interview location.

Interviews in Nigeria were conducted on a local university campus over a 3 week period. In Tanzania, the lead researcher recruited participants and conducted interviews in public settings over 6 weeks; this extension was needed due to election disturbances in October 2025, mandated curfews, and a country-wide Internet shutdown \cite{tz-internet}. The lead researcher in Zimbabwe conducted interviews over 1.5 weeks at a local university; this duration was accelerated in an effort to avoid recent and sudden international travel restrictions.

\subsection{Data analysis}
Data analysis proceeded in two phases: a deductive phase, where responses were categorized according to our research questions, followed by a thematic analysis to discover patterns within the interviews.
The lead researcher for Nigeria independently familiarized themselves with the 30~transcripts and participant data from Nigeria, as these were the first transcribed, then coded a random set of 10~transcripts to create a codebook capturing topics such as participant behavior, perceptions, and opinions on digital payment methods. The researcher then used this codebook to code the remaining 20 participants' data from Nigeria, revising and updating the codebook as necessary.

The leads from Tanzania and Zimbabwe independently corroborated the codebook by each coding the same random sample of three Nigerian transcripts. Disagreements were discussed and resolved to finalize the $67$ codes from Nigeria.
Once the remaining 60 Tanzanian and Zimbabwean interviews were transcribed, four researchers (the three country leads plus one additional researcher) coded the data. Interview questions and responses were divided into two roughly equal sections by length, each assigned to two researchers for independent coding. Researchers applied the finalized Nigerian codebook while inductively adding new codes to their local codebooks when they encountered themes not already captured.
This process grew the codebook from $67$ to $87$ codes, in two ways: some new codes captured themes specific to the Tanzanian and Zimbabwean data, while others came from decomposing existing codes into finer-grained ones—either because a code was too country-specific to apply consistently across all three contexts, or too broad to be applied reliably by different coders. The resulting codes were organized into five categories derived from our research questions.
All four researchers met over several meetings to discuss and resolve discrepancies.\footnote{Given our approach to analyzing a highly qualitative set of participant responses, a quantitative inter-coder reliability rating such as Krippendorff's $\alpha$ or Cohen's $\kappa$ was deemed unnecessary; McDonald \emph{et al.} note that such qualitative coding ``rarely, if ever, requires IRR''~\cite{McDonald:CSCW19}.} We iterated this process until all coding disagreements were resolved, achieving full agreement.
Our final codebook can be found in Appendix \ref{app:codebook}.

\subsection{Ethics}
This study was approved by our institutional review board (IRB), as well as the local ethics boards for each country.
Interviews were conducted in official local languages 
% English (Nigeria and Zimbabwe) and Swahili (Tanzania) 
for the clarity and comfort of participants. 
All participants provided informed verbal consent prior to participation, with consent procedures conducted in participants' preferred languages. 
Participants were compensated 
% (the local equivalent of) 10 USD, and reimbursed transportation costs up to 10 USD per participant, paid 
via mobile money, cash, or bank transfer depending on national norms and participant preferences. 
The compensation amounts were determined to be appropriate and non-coercive by our institutional IRB and the local ethics boards. 
We provide a more detailed overview of the ethical considerations and procedures of this study in Appendix \ref{sec:ethics}.

\subsection{Limitations}
First, as a qualitative study, our sample was limited to 90 participants (30 per country). The sample size and breadth of topics covered limit our ability to draw population-level conclusions or make statistical comparisons. However, our qualitative approach enabled us to examine participants' perceptions, experiences, and decision-making around digital payment systems in depth, including nuances that may be difficult to capture through quantitative measures alone.
Second, our sampling may have implicitly limited the diversity and representativeness of our participant pool. Our interviews focused on urban residents in each country, and snowball sampling may have further concentrated participants within similar demographic and social networks. Our eligibility criterion, which required participants to have used at least one category of digital payment system, may also have resulted in a sample with greater digital literacy or financial technology familiarity than the broader population.
To keep interview duration manageable and avoid participant fatigue, we could not probe every payment category with equal depth across all three research questions, and our interview protocol did not explicitly ask participants about digital payment systems they had chosen \emph{not} to adopt or use. This would have been useful for better understanding the relative lack of adoption of CBDCs, for instance. Nevertheless, several participants spontaneously discussed payment systems they chose not to use, providing useful evidence about barriers to adoption.

%% file: Sections/04-Results.tex
% Results
\section{Findings}
We present our main findings on drivers of adoption (RQ1, \Cref{sec:results-rq1}),  drivers of continued use (RQ2, \Cref{sec:results-rq2}), and obstacles to continued use (RQ3, \Cref{sec:results-rq3}). 
The most prevalent themes in our work were consistent across all three countries. However, some country-specific differences 
% observed in CBDC adoption and conventional bank perceptions, 
are outlined  in \Cref{sec:country-differences}. All other findings are presented jointly for the three countries and with nation-level distinctions noted where needed. We quote interview participants,  referring to them by participant ID, with Nigerian participants indicated by the NP prefix, Tanzanian participants indicated by the TP prefix, and Zimbabwean participants indicated by the ZP prefix.
As our survey is qualitative, we do not claim statistical significance for these results; we report raw numbers of respondents for reference.

\begin{table*}[t]
\centering
% \vspace{4pt}
% \setlength{\tabcolsep}{4pt}
% \renewcommand{\arraystretch}{1.15}
% \footnotesize
\resizebox{\textwidth}{!}{%
\begin{tabular}{P{1.5cm} P{4.0cm} P{4.0cm} P{4.0cm} P{4.0cm} P{4.0cm}}
\toprule
\rowcolor{white}
\textbf{} &
\cellcolor{headerMM}\textcolor{white}{\textbf{Mobile Money}} &
\cellcolor{headerMTA}\textcolor{white}{\textbf{Money Transfer Apps}} &
\cellcolor{headerCryptocurrency}\textcolor{white}{\textbf{Cryptocurrency}} &
\cellcolor{headerStable}\textcolor{white}{\textbf{Stablecoin}} &
\cellcolor{headerCBDC}\textcolor{white}{\textbf{CBDC}} \\
\midrule
\rowcolor{rowlabel}
\textbf{Common use cases and modalities}
&
\cellcolor{colMM}%
\begin{itemize}[leftmargin=*,nosep,topsep=2pt,itemsep=1pt]
  \item \textbf{Bill payments} --- utilities (data, rent), school fees
  \item \textbf{Merchant transactions} --- paying for goods and services in shops and receiving payment
  % \item \textbf{Small Businesses} --- receiving money through independent business ventures
\end{itemize}
&
\cellcolor{colMTA}%
\begin{itemize}[leftmargin=*,nosep,topsep=2pt,itemsep=1pt]
  \item \textbf{International transfers} --- remittances and receiving money from abroad
  \item \textbf{Business payroll} --- salary disbursement
\end{itemize}
&
\cellcolor{colCryptocurrency}%
\begin{itemize}[leftmargin=*,nosep,topsep=2pt,itemsep=1pt]
  \item \textbf{Investment} --- buy, hold for appreciation
  \item \textbf{Cross-border payments} --- cheaper transfers/no fees
  \item \textbf{Freelance/gig pay} --- remote workers paid in crypto, sometimes from international customers % where applicable
\end{itemize}
&
\cellcolor{colStable}%
\begin{itemize}[leftmargin=*,nosep,topsep=2pt,itemsep=1pt]
  \item \textbf{Savings} --- stability allows for reliable saving methods
  \item \textbf{Cross-border payments} --- cheaper transfers/no fees
  \item \textbf{Freelance/gig pay} --- remote workers paid in USDC/USDT
\end{itemize}
&
\cellcolor{colCBDC}%
None observed, as Tanzania does not have a CBDC \\
\\
\midrule
\rowcolor{rowlabel}
\textbf{Country-specific use cases and modalities}
&
\cellcolor{colMM}%
\begin{itemize}[leftmargin=*,nosep,topsep=2pt,itemsep=1pt]
  \item \textbf{Nigeria}: Mobile Money dominates everyday transactions due to distrust in commercial banking systems (improper management, lack of recourse, agent scams)
\end{itemize}
&
\cellcolor{colMTA}%
\begin{itemize}[leftmargin=*,nosep,topsep=2pt,itemsep=1pt]
  \item \textbf{Tanzania}: Commercial banking apps used more frequently locally than mobile money due to excessive fees
\end{itemize}
&
\cellcolor{colCryptocurrency}%
\begin{itemize}[leftmargin=*,nosep,topsep=2pt,itemsep=1pt]
  \item \textbf{Nigeria}: Expanded peer-to-peer (P2P) trading  (e.g., via Binance P2P, WhatsApp groups) after 2021 CBN ban on bank-facilitated crypto transactions  
  \item \textbf{Nigeria}: Crypto used for legally-questionable transactions due to low traceability
\end{itemize}
&
\cellcolor{colStable}%
% \begin{itemize}[leftmargin=*,nosep,topsep=2pt,itemsep=1pt]
%   \item \textbf{Nigeria}: USDT used as inflation hedge vs.\ naira volatility
% \end{itemize}
None observed
&
\cellcolor{colCBDC}%
\begin{itemize}[leftmargin=*,nosep,topsep=2pt,itemsep=1pt]
  \item \textbf{Nigeria}: Minimal eNaira (2021) uptake due to lack of public awareness and necessity
  \item \textbf{Zimbabwe}: ZiG (2025) adoption driven by favorable rates when used to pay for government goods and services
  % school fees, salary distribution, larger companies (supermarkets)
  % \item Tanzania: No CBDC currently established
\end{itemize}
\\
\bottomrule
\end{tabular}
}% end resizebox

\caption{Common and country-specific use cases and modalities observed, split by payment type. Bullets reflect themes drawn from our qualitative codebook; the country-specific row captures use cases and modalities observed in only one country's fieldwork interviews (these themes may exist in other countries, but they did not arise in our interviews).}
\label{tab:payment-use-cases}
\end{table*}

\subsection{Factors impacting initial adoption (RQ1)}
\label{sec:results-rq1}
Initial adoption marks the critical transition by the user to first trial or registration with a digital payment system. 
% Across our  interviews,  
We noted three common classes of factors impacting users' willingness to adopt a new digital payment system: (1) introduction paths, (2) promotions/features, and (3) international connections.

\subsubsection{Introduction paths and incentives}
% Participants learned about digital payment systems through various channels. 
% There are various avenues through which participants first learned about or were introduced to a digital payment system. 
Across all three countries, users learned about digital payment systems through interpersonal and community channels  more often than formal or mass-media sources; 
this reflects the central role of social networks and social influence in driving adoption of digital payment systems in these countries.

Peer recommendation was the most frequently reported pathway ($n=82$), with friends, family members, or acquaintances serving as primary introducers. Participants often described direct social influence that prompted them to adopt or start using digital payment systems. For instance, 

\begin{quote}
%ZP10: ``That one was my friends. Yeah, they were pressurizing me to get an Econet line because they were like, how do we send you money or how do we contact you? So I ended up buying the Econet line and then registering to EcoCash because I need money from them as well.''
ZP10: ``That one was my friends. [...] they were pressurizing me to get an Econet line because they were like, how do we send you money or how do we contact you? So I ended up buying the Econet line and then registering to EcoCash because I need money from them as well.''
\end{quote}

Brand reputation ($n=62$) ranked second, and encompassed trust transferred from well-known parent companies or ecosystem providers like MTN, Econet, and Airtel. Participants preferred adopting platforms backed by entities with established track records, viewing prior success as a signal of reliability.
% \begin{quote}
% %NP5: ``Yeah, I want to try this second one because they have a full history. Yeah, for those who deliver you just trust them, like okay just keep believing in them... MTN has done their own, MoMo, and it's still functioning till now because they have been records, successful records, proving records.''
% NP5: ``I want to try this second one because they have a full history. [...] those who deliver you just trust them, [...] just keep believing in them... MTN has done their own, MoMo, and it's still functioning till now because they have been records, successful records, proving records.''
% \end{quote}
% Public reputation ($n=47$) captured broader societal visibility and word-of-mouth perceptions, often reinforced by news reports, community discussions, or general awareness. A positive reputation often led to adoption, 
% e.g., 
% \begin{quote}
% ZP6: “I think reputation is important… if people had a good experience with you in the past, I'm likely to opt out \gf{opt out? Isn't that the opposite of what we want?} for that one, above anything else,''\end{quote} 
% while a negative one meant the user would shy away from adopting a particular payment system.
Public reputation ($n=47$) and advertisements ($n=32$) also influenced adoption rates positively, especially for mobile money and CBDCs. 
Participants reported encountering advertisements for digital payment systems through mainstream media channels such as television, radio, and infrastructural channels (e.g., telecommunications operator messaging).
% \begin{quote}
% TP14: ``I heard about mobile money through advertisements, on the phone, through messages that Vodacom sends, and on TV and radio.''
% \end{quote}
Other channels, including social media ($n=23$) and social influencers ($n=28$), were more prominent among younger participants in Nigeria and Tanzania, especially when first encountering cryptocurrencies or stablecoins. User 
%NP6 said, {``For Binance? I found out through... Twitter and a friend of mine. So more like I saw this. `Oh, we are doing, we are trading this, we are doing that.' And I saw it on Twitter.''} 
NP6 said, {``Binance? I found out through... Twitter and a friend of mine.''} 
% \nc{the rest of the quote was fairly redundant, so I omitted it.}
% And TP13: "I learned about these services through social media, and because they are systems that exist in our society and they are the methods we use, so I learned through social media systems.'
 % \end{quote}
Onboarding incentives (e.g. referral bonuses) also influenced adoption ($n=30$). 
% Some participants adopted Palmpay due to referral incentives:
% \begin{quote}
% %NP2: ``It was funny because Palmpay was doing something that if you like bring somebody, they'll give you like 500 (naira)... So that what he did, he sent me the link so I downloaded it.''
% NP2: ``Palmpay was doing something that if you [...] bring somebody, they'll give you like 500 (naira)... So that what he did, he sent me the link so I downloaded it.''
% \end{quote}

\subsubsection{Lack of choice or alternative options}
Many participants described encounters where they lacked choice ($n=52$) in using digital payment systems,
% This lack of choice was frequently tied 
due to policy interventions, cash scarcity, and related pressures. 
% that made digital alternatives unavoidable. 
In Zimbabwe, some participants ($n=10$) reported that they had been introduced to the CBDC by the government through salary payments. 
Individuals were required to open ZiG Bank accounts, in which part of their salaries could be deposited as ZiG, with no alternative option:
\begin{quote}
ZP30: ``So I work as a teacher and some money, um, we pay, we are paid through ZiG. So that's when I started using ZiG.''
\end{quote}
Participants in Nigeria and Zimbabwe cited the scarcity of cash, which forced them to use digital payment systems as cash substitutes. ZP11 mentioned,
\begin{quote}
``Okay, so basically, given the current situation in Zimbabwe, to actually find the physical ZiG currency is actually quite difficult.  
%\gf{is there a cash version of ZiG? I thought it was just digital}\nc{So, ZiG is actually the official currency now; the digital version of ZiG has been renamed GBDT according to Wikipedia. We may want to discuss this early on in the Background section \url{https://en.wikipedia.org/wiki/Zimbabwean_ZiG} -- Takudzwa please correct me if I am wrong}
It's like mining for gold.'' %\nc{that's insightful because the physical currency is literally supposed to be backed by gold}'
\end{quote}
Cash scarcity in both countries could have been exacerbated by policy frameworks. The Nigerian 
% In Nigeria, the 
2022–2023 cashless policy 
% imposed withdrawal restrictions, 
prompted many participants to adopt digital platforms to access or transfer funds without physical cash constraints \cite{onuegbu2025communicationawarenessacceptancedigital,ProtestsCashShortage}. 
Following the post-2024 launch of the Zimbabwean ZiG, persistent physical scarcity of both the newly introduced ZiG notes and the US dollar prompted users to adopt digital options
% , as cash availability remained unreliable 
\cite{CashShortageHits,einisCashScarceZimbabwe2023}.
%\gf{citation needed. TK can you find one?}.

% In Tanzania, participants used digital payment systems primarily for convenience rather than due to acute cash unavailability. Participants often cited reduced travel time, easier remote transactions, and widespread agent networks as practical incentives for initial adoption.
% \begin{quote}
% TP17: "This also makes it easier for me in my use, because there are times you cannot walk around with cash."\end{quote}

\subsubsection{(Dis-)Trust in government}
Paradoxically, both trust ($n=54$) and distrust ($n=42$) in government motivated initial adoption. Official endorsement such as regulatory approval, integration into national systems, or direct issuance e.g., eNaira in Nigeria or ZiG in Zimbabwe signaled legitimacy, scam protection, and reliability for many participants, prompting adoption.
\begin{quote}
NP8 : ``Since the government introduced eNaira, I felt it must be safer than these private apps that can disappear.'' 
\end{quote}
Similarly, ZP2 stated, ``I would choose the government [payment system], because if the government has accepted it, that means it has gone through processes, its legitimacy has been verified, and protections have been put in place for users.''
Conversely, distrust in government capacity, e.g., infrastructure reliability, privacy handling, or long-term stability, drove others toward private or decentralized alternatives at the outset.

 \begin{quote}
TP14: ``I think government has a lot of setbacks compared to other non-government companies, because private companies mostly try to satisfy customer needs.''
\end{quote}
This tension appeared across all three countries, though expressed differently (see \Cref{sec:country-differences} for an expanded discussion). Nigerian participants linked government distrust to historical policy volatility, Zimbabweans to currency instability, and Tanzanians to pragmatic comparisons favoring established mobile money or private platforms.

\subsubsection{International connection}
Global connectivity requirements and economic pressures drove initial adoption of cryptocurrencies and stablecoins. These factors were less prominent for %domestic-focused 
systems like MoMo or banking apps, but emerged strongly where participants needed alternatives to traditional cross-border money transfer platforms.

Cross-border payment capability was the most frequently cited international driver ($n=61$), with participants describing stablecoins, cryptocurrencies, and money transfer platforms as enabling faster, cheaper, and more reliable transfers than legacy remittance services such as bank wires. This appealed especially to those receiving diaspora remittances, engaging in regional trade, or supporting family abroad, where high fees, delays, and intermediary routing created significant friction.
\begin{quote}
TP16: ``So you can send money to someone in China, the United States, Japan, Germany, basically anywhere in the world, at any time, and it costs almost nothing, sometimes zero cost.''
\end{quote}
Notably, facilitating cross-border payments was one of the reported design goals for certain cryptocurrencies (e.g., Ripple \cite{qiu2019ripple}). 
Moreover, several users adopted stablecoins and cryptocurrencies as a hedge against inflation ($n=15$). 
% in contexts of currency volatility and depreciation. 
Participants viewed internationally-accepted, dollar-pegged assets like USDT as a way to preserve value amid local currency erosion, particularly when (fiat) USD access is restricted. 
% Some users  cited naira 
% % \gf{naira is not capitalized, I think, like dollar} 
% depreciation as a reason to move toward stablecoins %\nc{USD-pegged Stablecoins? Or crypto in general???} \gf{I don't think we had any mentions of non-USD-pegged stablecoins, so maybe we shouldn't make a distinction here...}]
% as a savings or hedging tool. 
% % \gf{Did users say this explicitly? If so, let's rephrase as "Nigerian users in particular cited persistent Naira depreciation..."}\nc{Yes, I did}
\begin{quote}
NP28: ``Well, another advantage I know was when the naira was still doing poorly against the dollar and euro behind the naira though I'm not seeing the effect. So having such currencies makes me defend against inflation as rising cost of goods.''
\end{quote}

\subsection{Factors driving continued use (RQ2)}
\label{sec:results-rq2}
% Our results identify the factors influencing users’ willingness to continue using digital payment systems across different use cases. 
We noted two main categories of factors influencing users' continued use of a digital payment system once they adopted it: (1)  usability considerations, and (2) security, privacy, and trust considerations. 

\subsubsection{Features of digital payment systems}
\label{sec:features-dps}
Continued use of digital payment systems was strongly influenced by practical features that supported everyday transactions, such as 
% Participants emphasized 
reliability, affordability, and ease of use.
% as central to their continued engagement. 
% whether a system remained usable over time.

\paragraph{Reliability, recourse and support.}
For many participants, continued use depended less on abstract trust and more on whether payment systems functioned reliably in everyday transactions. 
Participants favored systems that transferred funds quickly and without failure (reliability: $n = 57$; speed: $n = 67$) and that had demonstrated consistent performance over time ($n = 42$). 
% The ability to recover funds after errors was also particularly influential, with 

Participants described recourse mechanisms ($n = 77$) and support infrastructure ($n=57$) as central to deciding whether they would reuse a service. In Zimbabwe, ZP$2$ explained that, 
% recovering money after entering incorrect details directly increased trust:
% \begin{quote}
    ``...the fact that they were able to recover the money...makes them trustworthy. So I would use that service again.''
% \end{quote}
% In Nigeria, NP$22$ emphasized reliability through repeated successful use, explaining that:
% \begin{quote}
    % ``...is fast and reliable to an extent...I've not transferred to someone before, and they didn't get their money.''
% \end{quote}
% Support infrastructure influenced perceptions of dependability. 
Access to responsive customer service and clear resolution pathways influenced whether participants felt comfortable continuing to use a platform.
% (customer care accessibility: n = 57; responsiveness: n = 62). 
% In Tanzania, TP$21$ described trusting systems that corrected mistakes by reimbursing funds within a defined timeframe:
% \begin{quote}
%     ``if at all it occurs that either the system is down...or you make a mistake in sending...then the money is reimbursed back after 24 hours.''
% \end{quote}

Finally, physical points of contact strengthened perceptions of reliability, particularly when digital channels failed ($n = 57$). Participants described in-person branches as providing reassurance that problems could be addressed directly. Together, these accounts show that continued use was closely tied to systems that not only worked efficiently but also offered dependable mechanisms for recovery and support.

\paragraph{Convenience, cost and accessibility.}
Everyday convenience played a central role in determining which digital payment systems participants continued to use, with a particular emphasis on time savings ($n=68$). 
% Systems that reduced effort and saved time were consistently preferred (n = 68). 
% In Nigeria, NP$23$ explained about digital payments:
% \begin{quote}
%     ``It saves time a lot. It saves the time you go to the bank to queue to fill a teller to either withdraw or send your money.''
% \end{quote}
Convenience was enhanced by geographical accessibility---participants relied on systems that were more widely accessible and accepted across locations ($n = 61$)---and interoperability with other digital payment options. In Tanzania, TP$1$ described mobile money as easy to use because services were available, stating, 
    ``it is easily available, and everywhere I go is there, I can scan and pay.''
Access was enhanced by alternative channels that did not require internet connectivity ($n = 29$). Meanwhile, TP$19$ stated his preference for \textit{Binance} stemmed form its interoperability with mobile money: 
\begin{quote}
    ``Because Binance is also connected to these M-PESA, TIGOPESA, AIRTEL MONEY, they are connected.''
\end{quote}
% Broad accessibility reinforced convenience. 
% Affordability was another major factor supporting continued use. 
Participants also favored platforms with low or minimal transaction fees, particularly for frequent transfers ($n = 57$). 
% ZP$20$ described choosing a service because its fees were not excessive, stating, ``I usually use EcoCash because even their charges are not that extreme.''
% \begin{quote}
    % I usually use EcoCash because even their charges are not that extreme.
% \end{quote}
A smaller number of participants also emphasized charge transparency, noting that clearly explained deductions increased confidence ($n = 5$, more discussion in \Cref{sec:undesirable-features}).
% Participants described USSD options as enabling transactions even in low-connectivity settings.
% \gf{why is this? USSD goes over the carrier network, no?}

%Finally, participants described relying on systems that were widely accessible and accepted across locations and contexts ($n = 61$). In Tanzania, TP$1$  described mobile money as easy to use because services were available, stating, 
%    ``it is easily available, and everywhere I go is there, I can scan and pay.''
%Access was enhanced by alternative channels that did not require internet connectivity ($n = 29$). 

These accounts indicate that ease of use, manageable costs, and accessibility across environments were key to sustaining everyday engagement with digital payment systems.

% Overall, continued engagement reflected a combination of trust-related considerations and practical usability in everyday transactions.

\subsubsection{Security, privacy and trust}
\label{sec:sec-priv-rq2}
Across all three countries, participants described continued use as contingent on how trustworthy systems felt in practice. 
% based on visible legitimacy, perceived intervention capacity, and control over money and personal information.

\paragraph{Institutional and platform trust.}
Participants linked continued use to visible institutional legitimacy and platform authority ($n=38$). Systems backed by recognized regulators were viewed as more trustworthy, particularly when users believed these entities could provide accountability or recourse. NP$2$ noted that: 
% seeing central bank approval within an application increased confidence:

\begin{quote}
    ``If you enter Opay, hit ‘down’, you will see ‘approved by CBN [Central Bank of Nigeria]’. If I have problem, I can take it up like with CBN or to anybody.'' 
\end{quote} 
Similarly, in Zimbabwe, ZP$3$ perceived platforms operating under the Reserve Bank as legitimate. 
% having passed a legitimacy threshold. 
% stating that “passing through the RBZ” meant the platform had “gone through a certain vetting.''

Beyond formal regulation, participants emphasized trust in cryptocurrency and stablecoin platforms 
% rather than specific coins themselves. Trust was shaped by perceptions of the platform's legal recognition, 
based on global reputation ($n=17$).
% and ability to safeguard assets during transactions ($n=28$). 
% Participants described trusting popular cryptocurrency platforms that were perceived as  legally-recognized, independent of the particular coin being traded. 
TP$6$ noted:
\begin{quote}
    ``The platforms I use, especially Binance, have strong security...their institutions are somewhat known legally and recognized.''
\end{quote}
 % For some participants, 
 % % this trust extended beyond national boundaries, with 
 % global reputation served as a signal of reliability ($n=17$). 
 % In Zimbabwe, ZP$9$ expressed confidence in internationally-recognized services, noting, 
 % \begin{quote}
 %     ``It’s a world-recognized payment system, so I don’t think they will tamper with our information.''
 % \end{quote}
  
  Participants further associated trust with a platform’s capacity to act decisively during disputes ($n=25$). NP$6$ recounted a case where a complaint resulted in immediate account restriction, interpreting this response as evidence of effective enforcement: 
  % \begin{quote}
      ``I just launched a complaint, and immediately I think they blocked or restricted his account.''
  % \end{quote}
  
Together, these accounts suggest that institutional backing, global reputation, and platform intervention capabilities play a central role in sustaining users’  continued use.

\paragraph{Perceived risk.}
Visible security mechanisms that reduced the likelihood of unauthorized transactions played an important role in continued use ($n = 47$). In Tanzania, TP$13$ described confidence in systems that required multiple verification steps:

\begin{quote}
    ``I believe the money is kept safely because there are... 
    % verification methods. If you want to send to someone, there are v
    verification steps for sending. You can use passkeys or an authorization indicator to send to another person, so I believe it is safe.''
\end{quote}

Perceptions of financial risk were also closely tied to value stability, particularly in contexts marked by currency volatility ($n=30$). While institutional or government backing was often  a source of reassurance in stable monetary environments, this assurance weakened in settings with a history of inflation and currency depreciation. 
% \nc{This doesn't flow super well, because so far we have been saying that institutional/government backing reassures people, but here, it's the exact opposite. I think we probably need some sort of contrasting sentence here.} \tk{I have editted the paragraph}
In Zimbabwe, participants contrasted CBDC instability with alternatives perceived as more reliable. ZP$21$ explained:

\begin{quote}
    ``We can keep in dollars rather than ZiG because it keeps losing value.''
\end{quote}
Participants in Nigeria and Tanzania similarly described preferring U.S. dollar–pegged stablecoins %such as USDT 
to avoid sudden fluctuations and preserve value during transactions.
Overall, participants’ engagement with digital payment systems depended on whether perceived risk (both security risk and financial risk) aligned with their intended use and tolerance for oversight.

\paragraph{Privacy, surveillance and control.}
% Participants' engagement with digital payment systems depended on how much control they felt they retained over their personal information and transaction visibility. 
Some participants described privacy protections as supporting continued use, particularly when platforms limited the amount of personal data required ($n=38$). In Zimbabwe, ZP$13$ expressed confidence in systems that were perceived to operate under data protection requirements, explaining,
% \begin{quote}
    ``I read there is a data protection act in every organization where the data of the customer has to be protected.''
% \end{quote}

In contrast, other participants in Tanzania and Nigeria associated government-linked payment systems with heightened surveillance, which made them more cautious and selective in how they used these systems ($n =10$). 
% Perceived oversight did not necessarily discourage use, but influenced users' comfort around specific transactions. 
In Nigeria NP$5$ described concerns about transaction monitoring and potential account restrictions by authorities, noting that:
\begin{quote}
    ``CBN [Central Bank of Nigeria] can even track your records...you are doing some transaction, they can block your account and say an investigation is needed to find the source of money.'' 
\end{quote}
Participants in Tanzania similarly described discomfort with government-linked payment systems due to extensive transaction oversight, expressing concern about increased scrutiny over recipients and transaction purposes.
Some participants perceived decentralized systems as having reduced external visibility, which increased their trust in the system ($n=24$). 
% TP$16$ described feeling more comfortable using decentralized systems because of the absence of external intervention:
\begin{quote}
    TP$16$: ``These are decentralized systems. So I feel very comfortable when my money is there because I know no one can see it, hack it, or do anything to it.''
\end{quote}
A smaller set of participants ($n=2$) described using cryptocurrencies specifically for their anonymity and limited traceability, including their ability to facilitate fraud. In Nigeria, NP$30$ explained that these properties made cryptocurrencies appealing in legally ambiguous contexts:

\begin{quote}
    ``Based off fraud, that’s like the easiest and fastest way to get money... either Bitcoin or Ethereum.''
\end{quote}
Overall, preferences around privacy, surveillance and data control influenced how participants engaged with digital payment systems.

\subsection{Failures, risks and breakdowns (RQ3)}
\label{sec:results-rq3}
While \Cref{sec:results-rq2} shows that users generally continued using digital payment systems due to a combination of (1) desirable usability and financial features, and (2) security, privacy, and trust properties, we conversely found many users frustrated with the lack of these properties in other systems.

\subsubsection{Undesirable features}
\label{sec:undesirable-features}
 Our analysis revealed three primary (and sometimes overlapping) dimensions of feature-based challenges users face post-adoption: (1) high and opaque financial costs, (2) poor user experience, and (3) accountability concerns.

\paragraph{High and opaque financial costs} In all three countries, financial barriers including excessive charges ($n=51$) emerged as an important concern.
% being the dominant qualitative code. 
Participants reported high (and sometimes undisclosed) transaction fees across all categories of digital payment systems, especially in money transfer platforms and CBDCs. 
% When discussing Money Transfer platform fees, TP27 said:
% \begin{quote}
%     TP27: ``So the disadvantage is you only see the money you are sending, so there are charges you don’t know. You are not sure of their rates.'' 
%     % So they can charge you more and you don’t know.'' 
%     % And then the charges that I know, the charges are too high. 
% \end{quote} 

% ZP10 also recalls that with the ZiG:
% \begin{quote}
%     I think to a certain extent it's trustworthy, and then to a certain extent it's not, because of the charges that they take from every transaction that you do.
% \end{quote}

% This behavior appeared mainly in Nigeria and Tanzania, as Zimbabwe had low cryptocurrency and stable-coin adoption rates in our participant pool. 
% NP3 states:
% \begin{quote}
%     But on withdrawals they kind of have some huge percentage of charges on withdrawals, on USDT platforms
% \end{quote}

% Although mobile money applications as a whole were the least scrutinized when discussing excessive transaction charges, this observation varied based on participants' experience with specific mobile money companies. NP11 recalled:
% \begin{quote}
%     But Palmpay the reason why I stop using Palmpay for now is because of their charges.
% \end{quote}
% TP4 explained:
% \begin{quote}
%     The disadvantages, I would say high transaction fees, like in AzamPay, there's pretty much high transaction fees when you need to pay for a service or send money. 
% \end{quote}

Charges were not only incurred by transactions, but also periodic deductions from both idle and active accounts. These fees were variously labeled as ``account maintenance'', ``card maintenance'', or applied without justification. This was especially common with conventional banking applications, particularly among Nigerian participants. NP$24$ shared:
\begin{quote}
    ``They charge a lot even without making any transaction. You get debited for card maintenance. You get debited for whatever.'' 
    % Like the charges are too numerous.'' 
\end{quote}

Transaction fees depended on the transaction amount, the destination, and the number of transactions a participant made in a day. As a result, \textbf{many participants rotated digital payment systems or dropped them altogether} to save money ($n=44$). Those who adopted this strategy described maintaining multiple payment systems, coins, or currencies, and strategically switching between them to avoid excessive costs. 
% Incidentally, the contexts of this behavior varied across the three countries (Nigeria: $n=10$, Tanzania: $n=11$, Zimbabwe: $n=23$).

In Tanzania and Nigeria, participants most frequently switched payment methods to circumvent gas fees or navigate mobile money limitations and cross-platform incompatibilities or charges. 
% Cross-platform transactions, between different mobile money providers or from one service type to another, would incur extra charges. 
Daily transaction limits on free transfers also prompted some participants to switch to alternate mobile money accounts once they exhausted their free transactions. 
% TP6 said, 
% \begin{quote}
     % ``If you send using mobile money the charges are high, but if you send through the bank, especially bank-to-bank, the charges are low.''
% \end{quote}
NP13 said,
% \begin{quote}
%    ``So the situation was that I think I use my free transfer on Opay... I transferred it to Palmpay so that I can be able to use until the free transfer.'' 
    ``I use my free transfer on Opay [...] transferred it to Palmpay so that I can [...] use [...] the free transfer.'' 
% \end{quote}
% When avoiding excessive gas fees, participants either traded the unfavorably priced coin, or changed the platform they are trading on. NP7 stated: 
% \begin{quote}
%     I have tone so I would want to like change to the one that has the lowest gas fee most of the time.
% \end{quote}
In Zimbabwe, currency switching is primarily used to avoid unfavorable exchange rates. Due to ZiG volatility, government entities accept it at lower rates compared to the US dollar while many vendors prefer US dollars. 
% ZP3 explained:
% \begin{quote}
%     ``...into the shop, I cannot use the US dollar or the ZiG, because the rates in there, they are actually ridiculous. So I need to use the what? The US dollar compared to the ZiG.
% \end{quote}
% Conversly, 
ZP28 stated:
    ``I usually use this service to pay for government levies or fees because its cheaper to use [...] ZiG rather than [...] US dollars.''

High and unpredictable costs (including gas fees) were also cited by participants transacting cryptocurrencies and stablecoins. 
% who  complained about  fluctuations. 
% Gas fees were sometimes expected to drop and rise in a matter of hours, causing participants to wait out undesirable charges or act rashly. 
Cryptocurrency (non-stablecoin) volatility risk ($n=30$)  emerged as a major concern for current and former traders. Participants lost funds through price fluctuations or irretrievable seed phrase loss. While some accepted volatility as part of cryptocurrency trading, others abandoned it entirely after significant losses. NP17 shared:
\begin{quote}
    ``I lost over 200 dollar[s] [...] that [...] I wanted to use for us to pay our house rent. I cried.''
\end{quote}

% Lastly, "Transaction Limits" ($n=27$) financially affected participants across all three countries, primarily those using mobile money and CBDCs. Limits on transaction amounts, daily frequencies, and account balances forced users to alternative payment systems for large or frequent transfers. Discussing savings, TP6 explained:
% \begin{quote}
%     Because with mobile money and so on, there is a limit, like three million, five million. So storing a large amount above that is hard.
% \end{quote}

\paragraph{Poor user interfaces and experiences}
Participants across the three countries faced technical demands and usability challenges. Network challenges ($n=76$) were widely-cited, resulting in stalled transactions, debited but unreceived funds, or complete service unavailability. 
% Many prior negative experiences stemmed directly from network downtime and subsequent failure to return funds when lost. 
% During a time-sensitive situation, 
TP7 shared that:
\begin{quote}
    ``There was a time whereby they actually told us our services will be down today, but I had to pay my school fees, and the due date was really close, so that… that was really painful.''
\end{quote}

Inconvenient processes ($n=39$) 
% was the last notable challenge experienced by participants. These processes 
included the procedures required to utilize the digital payment system, incompatibility, and complexities with some recourse mechanisms. 
% When utilizing Money transfer services, 
ZP21 said,
% \begin{quote}
    ``But what I know about Western Union is that you can travel some distance and there are so many paperwork needed there.''
% \end{quote}

Many cryptocurrency and stablecoin users recognized that a cryptocurrency literacy requirement ($n=35$), including wallet address management ($n=25$), was essential. Users acquired knowledge through cryptocurrency social media groups, influencers, and online resources to ``predict'' markets. 
% NP19 articulated this necessity:
% \begin{quote}
%     The thing is like when you get into the crypto space, you need to be educated because for you to use the platform you need to have a degree of literacy about it.
% \end{quote}
% Basic understanding of how cryptocurrency and stablecoin wallet address management was also viewed as critical ($n=25$).
% , especially in the case of "Wallet Address Management" . 
Traders acknowledged the risks attached to seed phrase security and wallet address accuracy, exercising caution in these processes: 
% TP2 explained:
\begin{quote}
    TP2: ``As I said, it is risky. One mistake and you lose everything. If you enter the wrong link, you lose everything, and it is non-reversible.''
\end{quote}

\paragraph{Accountability concerns}
The final dimension centered on recourse availability, consistency, and responsiveness when funds were lost. The majority of cryptocurrency or stablecoin users  expected no remedy for lost funds  ($n=35$). When asked asked about anticipated help following fund loss, NP26 reiterated, ``You're in a helpless situation. There is no connection to call or anything.''
% \begin{quote}
%     You're in a helpless situation. You're in a helpless situation. There is no connection to call or anything.
% \end{quote}
%\nicolasc{this is not true for CBDC or is it?}

Even for payment systems like mobile money and CBDCs where recourse was available, users still faced challenges. Inadequate customer support ($n=31$) manifested as slow or absent responses and failure to return funds. 
TP23 explained, ``When you ask for help, they tell you to contact the recipient. Then they tell you to wait 72 hours for a refund. When the 72 hours pass and you ask again, they tell you that the money was already used.''
Interestingly, 
% "Recourse Dependent on Amount Lost" ($n=18$) emerged as 
participants reported different treatment based on loss magnitude ($n=18$); either companies dismissed small amounts or users determined recovery processes were not worthwhile for minor losses. 
% When asked whether users would reuse a service on which they previously lost money, some participants based their decision on the magnitude of the monetary loss.
% TP23 commented: 
% \begin{quote}
%     When you ask for help, they tell you to contact the recipient. Then they tell you to wait 72 hours for a refund. When the 72 hours pass and you ask again, they tell you that the money was already used.
% \end{quote}

\subsubsection{Security, privacy, and trust}

\paragraph{Security} Participants across the three countries expressed substantial security concerns, primarily manifesting as a fear of scams ($n=48$). 
% within the "Fear of Scams" code, totaling 48 participants. 
% Third-party scammers or bankruptcy 
Anxieties about third-party fraud and platform bankruptcy drew from personal experiences, news reports, and secondhand accounts. Commonly-cited scams included fraudulent telephone calls, cryptocurrency ``pump and dump'' schemes \cite{hamrick2021examination,xu2019anatomy}, Ponzi operations, and system collapses. NP17 said,
% \begin{quote}
    ``Even some Nigerians were doing fake coins... 
    % If you're not, 
    if one is not careful, and you fall into that trap.''
% \end{quote}
Zimbabwean participants frequently referenced "E-Creator," a notorious Ponzi scheme that defrauded thousands of users of over 1 million USD within months, building trust with celebrity endorsements, physical offices in Zimbabwe, and integration with other popular platforms like EcoCash \cite{e-creator}. 
% This scam gathered users and garnered trust through celebrity endorsements, physical offices in Zimbabwe, and integration with other popular platforms like EcoCash. When users began reporting problems accessing funds, the platform abruptly shut down, taking its users' funds with it. 
% ZP29 recounted: 
% \begin{quote}
%     ``We have, it's, the e-creator, as I was… as I alluded to earlier, yes, it was new, and people liked to join, and they were scammed from it.''
% \end{quote}

Scams were not only a digital affair, but also a physical one:  reportedly ($n=9$),
agents and bank clerks could scam users out of their money under the guise of assisting them.
This contrasts with the cited benefits of human agents for recourse and convenience (\Cref{sec:features-dps}).
% While many users, as described in previous sections, appreciated the accessibility of live and in-person assistance with their digital payment service, some were defrauded by those same agents. 
% experienced situations where that trust was breached for selfish reasons. 
ZP7 explains:
\begin{quote}
    % ``they make it hard for\ldots me to trust them, because I've\ldots  
    ``I've had occurrences whereby\ldots  their agents would give out fake money, so that caused me not to actually have any trust in them whatsoever.''
\end{quote}

Fear of hackers was prevalent ($n=27$).
% gaining unauthorized access to funds and personal information also plagued participants' security concerns ($n=27$). 
Drawing from news reports and secondhand accounts, this anxiety extended across all digital payment system categories. TP4 said,
% \begin{quote}
    ``I heard different cyber attacks that the system faced and how people lost money from using this payment system, so when you get information like that, you don't expect me to trust them.''
% \end{quote}
% These concerns underscore the importance of security in digital payment systems. 
% , digital fraud, platform collapses, and in-person threats, and hackers, underscore how trust in digital payment systems extends beyond technical reliability to even threats outside of the system. 
Accumulated knowledge from personal losses, community experiences, and public news strongly shaped participant behavior, and a single negative experience could impact people's willingness to use a product.
Our finding parallels prior literature showing the negative impacts of security incidents on people's perceptions and behaviors \cite{rebensky2021user,de2023analysis}.
% This suggests that security fears could represent a fundamental barrier to increasing digital payment engagement, with users maintaining skepticism even toward systems they regularly use.

\paragraph{Privacy} Participants grappled with trust in data handling practices. Participants  distrusted centralized digital payment systems with their personal information ($n=35$), fearing their  information would be sold, shared with third parties, or given to the government. 
% ZP18 shared: 
% \begin{quote}
%     ``On the personal information, I'm not 100\%\ sure to be honest. Because, [...] they probably have third party vendors that they might share your information with.''
% \end{quote}
Cryptocurrency and stablecoins were not mentioned among these concerns, possibly due to the lack of a central authority and relatively minimal collection of personal information. 
% required to open an account. 

% Some privacy concerns stemmed from unawareness of platform issuers ($n=18$), 
% including uncertainty about who operated platforms and how they managed user data. 
% While this knowledge gap did not always erode trust, 
% participants mentioned being unable to fully trust systems with unclear operator and data practices. 
% NP7 said:
% \begin{quote}
%  How trustworthy are they? I really can't. I really don't know. And the reason being that is that a lot of us that we like say we trade in USDT, USDC, we don't really even know which company is behind it.
% \end{quote}

\paragraph{Trust and the interplay between platform categories}
Participants maintained nuanced and varied trust levels in different payment system categories. 
Many users allocated funds according to their trust assessments, rather than fully committing to any single type of platform  ($n=31$), often reserving large funds for more established systems such as conventional banks.  
% These assessments manifested in 
% distinctions are made based on 
% the amount a user chose to  keep in a given account, often reserving large funds for more established systems such as conventional banks. 
% This observation was most prominent in Nigeria ($n=18$), then Tanzania ($n=10$) and ($n=3$) from Zimbabwe. 
TP27 explained,
% \begin{quote}
    ``I cannot keep a big amount [...] on mobile money, but in the bank I can [...].''
% \end{quote}

Several users ($n=16$) expressed distrust in peer-to-peer cryptocurrency exchanges. 
Nigerian and Tanzanian participants reported inconsistent exchange rates and incomplete transactions when trading with third-party dealers to convert cryptocurrency to local currency. 
NP2 even reported getting into a physical altercation with a trader who failed to uphold their side of the trade. 
Indeed, peer-to-peer cryptocurrency exchanges have been shown to fail at preventing scams, and are extensively used in a number of countries including Nigeria \cite{Tsuchiya:WWW24}.

% When trying to retrieve their funds, NP2 stated:
% \begin{quote}
%     "I was looking for like somebody to help me to convert it. So I gave it someone else to help me, but the person ended up blocking me this and that.
% \end{quote}

% Security and privacy risks come at a high cost and serve as significant barriers for users navigating this vast digital landscape. Participants' fears of individual and company-wide scammers, malicious online actors, and undisclosed data practices were grounded in real-life events they directly experienced or witnessed within their communities. 

% In response, participants exhibited high levels of financial literacy by allocating funds to different services according to their trust levels, mitigating concentrated risk exposure. 

Despite diversifying their holdings in different categories of accounts, participants still operated largely under an umbrella of suspicion due to unawareness of platform operators and data handling practices. 
% This fragmented trust suggests that security and privacy failures constrain how deeply users can integrate digital payment systems into their financial lives. 
% possibly limiting the potential of these technologies.

%% file: Sections/05-Discussion.tex
% Discussion
\section{Discussion}
We next discuss implications of our results, highlighting first nation-level differences, before summarizing these results in a concrete set of takeaway messages from which we derive recommendations.
\subsection{Nation-level differences}
\label{sec:country-differences}
While most of our findings were consistent across countries, we did note several  country-specific behaviors.
Here, we list them and provide relevant national context.
% to help explain some of these differences. 
% , which help inform our future recommendations. 
% The distinct digital payment system adoption and usage behaviors across Nigeria, Tanzania, and Zimbabwe reveal how national contexts deeply affect user experiences and perceptions. 

% \textbf{CBDC} The stark contrast between e-Naira's failure ($n=2$) and ZiG's relatively higher engagement ($n=26$) illustrates how institutional trust shapes CBDC adoption, even in the presence of ZiG's fluctuation hurdles. 

\paragraph{A tale of two CBDCs}
The two surveyed countries with a CBDC (Nigeria and Zimbabwe) highlighted starkly different stories. 
% which appeared correlated with government involvement and financial factors. 
% When interviewing the two participants who adopted e-Naira, 
Our (very limited) pool of eNaira users ($n=2$) reported limited platform usage, primarily due to minimal public awareness and adoption, as well as distrust in the government. 
While our sample size is too small to generalize, adoption data shows that  eNaira wallet downloads reached only $0.8\%\ $ of bank accounts, with 98\%\ of wallets remaining inactive by mid-2023 \cite{imf2024cbdc, wezel2023nigeria}.
% NP22 described e-Naira as an unnecessary intermediary step: withdrawing funds from conventional banks, depositing into e-Naira, then transferring to whichever mobile money platform the recipient used. Specifically, 
\begin{quote}
    NP22: ``I think it later phased out... a lot of people didn't know about it....I would have given it a chance...  if everyone was using it.''
\end{quote}
Our Nigerian subjects reported seeing no compelling reason to adopt a government-backed digital currency given governmental distrust, especially while other new and reliable mobile money alternatives were available (i.e. Palmpay, OPay). 
% \gf{Isabel, are u sure we should say this ("institution itself lacked credibility")? What were the reliable alternatives?}
% \begin{quote}
%     I think it later phased out and I didn't exactly use it like over and over again... because a lot of people didn't know about it....Yeah, because everybody would be using Enaira to Enaira. I would have given it a chance like a better chance at exploring it and knowing how it actually worked if everyone was using it."
% \end{quote}

% Nigeria's e-Naira suffered virtually no uptake due to government distrust rooted in corruption, administrative negligence, and lack of incentives.    

On the other hand, Zimbabwe's ZiG saw higher adoption despite currency volatility, possibly due to cheaper rates and heavy government 
influence; in fact, ZiG effectively replaced the Zimbabwean dollar in 2024. 
% We also observed somewhat higher government trust in Zimbabwe ($n=21$) than in Nigeria ($n=14$), though we stress that we cannot draw conclusions from these numbers alone. 
% Our Zimbabwean interviewees 
Participants recalled hearing about the CBDC launch directly from the administration, and those working in government or public education were forced to receive their salaries in ZiG. 
\begin{quote}
    ZP30: ``So I work as a teacher and [...] we are paid [in] %through with the 
    ZiG. So that's when I started using ZiG.''
\end{quote}
Additionally, Zimbabwean distrust in government centered on economic volatility rather than broad distrust in the government, representing a fundamentally different type of distrust than that seen in Nigeria. 
Studies in the Netherlands \cite{bijlsma2021triggers} and globally \cite{ngo2023governance} have similarly shown that government trust drives positive sentiment about CBDCs, 
though these studies saw more trust in the government.
% , albeit with higher overall levels of trust in government.
% from Nigeria's institutional rejection. 

% Contrastingly, ZiG launched at a time of monetary instability without fresh competition in the digital finance market. 

\paragraph{Timing strongly affects product perception} 
A second source of nation-level differences was the timeline on which services were released; the case of mobile money is particularly informative. 
% The timeline of mobile money introduction altered how users in each country perceive conventional banking and alternative payment systems. 
In Nigeria, persistent conventional banking failures, including transaction problems ($n=28$) and inadequate resolution mechanisms ($n=18$) decreased trust in these systems. 
% NP29 lamented:
% \begin{quote}
%     First bank have frustrated me a lot. Just…. I told you the story of someone sent me money. I've not seen the money and the money hang, two days I slept in Ebonyi. There's money in my account. I can't withdraw. The man kept calling me. I said I've sent this money. I went to the bank. The bank said that yes, I have transferred the money. But the money  hanged...This thing I'm telling you up till the man gets me arrested. 
% \end{quote}
The relatively recent launch of mobile money services like OPay in 2018 and Palmpay in 2019 \cite{opay,PalmPay} disrupted this pattern and provided users with a superior alternative. The failure of conventional banks and the 2020 Cashless Policy \cite{yaqub2013cashless} positioned mobile money  uniquely  for widespread success. 
In contrast, Tanzania and Zimbabwe introduced mobile money services nearly a decade earlier (M-Pesa in 2008 and EcoCash in 2011, respectively), concurrent with conventional banking app development. 
% This parallel evolution meant 
Users in these countries thus perceived mobile money and conventional banking as equivalent digital options, rather than competing substitutes. 

\subsection{Theoretical implications}
% \gf{Karen added this section, pls review}
While our findings provide empirical insights into the adoption and use of digital payment systems across three African countries, they also have broader implications for existing theories of technology adoption, trust, and continued use. Rather than proposing a new theoretical framework, we use established theories to interpret our findings and highlight where our results extend, refine, or challenge existing perspectives on digital payment systems.
Our findings bridge prior work on voluntary models of adoption with recent studies exploring less voluntary reasons for adoption.

\paragraph{
% Extending technology adoption theories \ks{
Expanding conceptions of technology adoption from choice to 
necessity} Technology adoption models such as Technology Adoption Model (TAM) and Unified Theory of Acceptance and Use of Technology (UTAUT)  primarily explain adoption through individual perceptions, including usefulness, ease of use, social influence, and facilitating conditions \cite{venkatesh2003user,davis1989technology}. Consistent with these models, our findings show that peer recommendations, brand reputation, and perceived convenience influenced adoption. However, participants also described adopting digital payment systems because of structural conditions that include institutional mandates, cash scarcity, monetary instability, and cross-border economic needs---conditions that could substantially outweigh voluntary evaluation of technological characteristics. These findings suggest that digital payment adoption in emerging economies is jointly shaped by individual perceptions and broader institutional-economic contexts. 
Similar results have been found in China and India, where mobile payments gradually became infrastructural; as this transition occurred, many vendors felt obligated to adopt to participate in the economy \cite{shen2020can,he2023have,Pal2018}. 
% a similar trend resulted after India's demonetization \cite{Pal2018}.
% \gf{Can we find a few other examples of people feeling forced to adopt fintech?} \ks{
% Similarly, He et al. found that senior street vendors in China passively adopted mobile payments in response to business requirements, while Pal et al. found that cash shortages following India's demonetization pushed merchants toward digital payments despite reservations about their use \cite{he2023have,Pal2018}. 
Together, these findings suggest that digital payment adoption is shaped not only by individual perceptions of a technology, but also by institutional and economic conditions that can constrain whether adoption is meaningfully voluntary.

\paragraph{Expanding conceptions of trust in digital payment systems} While the preceding discussion broadens the conditions under which technology adoption occurs, our findings also broaden the types of trust considerations that shape adoption and continued use. Prior work on digital payment adoption globally commonly conceptualizes trust as an antecedent to technology adoption and continued use, focusing primarily on trust in the technology or service provider 
% (\ia{\cite{sowon2024role, 10.1145/3411764.3445065}})
% \gf{Update these citations, they are not a perfect fit}
% \gf{Suggested alternative citations: 
\cite{zhou2011effect,hillman2014user, chandra2010evaluating, xin2015antecedents}. 
HCI research has similarly highlighted the centrality and complexity of trust in emerging financial technologies \cite{sas2015exploring}.
Our findings suggest that participants evaluated multiple, 
% \gf{an even broader number of(?)}, \ks{and 
even broader, dimensions of trust, including technical trust (system reliability and security), institutional trust (confidence in governments, regulators, providers), economic trust (financial stability), and informational trust (handling of personal information). Rather than evaluating these dimensions independently, participants navigated trade-offs between them, e.g., preferring technically reliable platforms despite privacy concerns or selecting stablecoins for economic stability despite lower institutional oversight. Informational trust, in particular, is well explained through the lens of contextual integrity \cite{nissenbaum2004privacy,Tronnier2024}. %\cite{nissenbaum2004privacy,Tronnier2024, Kadiri2025-KADCII-2, 10.1145/3313831.3376415}. 
Participants rarely objected to information collection itself; rather, they evaluated whether information sharing aligned with the expected roles of governments, payment providers, and third parties. Information sharing for transaction processing, fraud prevention, dispute resolution, or regulatory compliance was generally viewed as legitimate,
% and, in many cases, strengthened trust.  In contrast, 
while data sharing to facilitate surveillance or third-party sharing was generally not. 
% or interventions beyond the expected role of the recipient. T
These findings add nuance to the multidimensional scope of trust in digital payment systems. 
% with informational trust depending not solely on security or privacy mechanisms, but also on whether information flows conform to users' context-dependent expectations regarding who should access their financial information, under what circumstances, and for what purpose.

\paragraph{Continued use as a dynamic allocation of financial activity across payment systems} %Existing digital payment literature examines the adoption or continued use of individual payment technologies (\ia{\cite{Pal2018, Lewis2019}}).
Prior work across mobile payments, mobile money, cryptocurrencies, and other digital payment technologies has largely conceptualized continued use with respect to a focal technology or service, examining factors that determine whether users intend to continue using that system \cite{zhou2013empirical, chen2017understanding, odoom2020mobile, alhassan2020consumer, al2024people,he2025antecedents}. In contrast, HCI research on everyday financial practices has shown that people manage financial activity through broader ecologies of payment technologies and services. Our findings connect these perspectives by showing that continued use within such an ecology is not simply a binary state of continued use or abandonment, but involves the dynamic allocation of financial activity across available payment systems \cite{perry2018moneywork,Lewis2019,kotut2024things}. %Our findings} Our findings instead suggest that continued use is an ongoing process of negotiating trade-offs across multiple digital payment systems.
Participants routinely switched between banks, mobile money, cryptocurrencies, stablecoins, and CBDCs balancing factors such as  transaction fees, exchange rates, accountability mechanisms, privacy expectations, and perceived risks. Negative experiences—outages, scams, hidden fees, or failures of recourse—did not necessarily lead to abandonment,  but prompted redistribution of financial activity across platforms, choosing different systems for different purposes. These findings suggest that continued use is better understood as dynamic management of a fragmented payment ecosystem than as persistent commitment to technologies.

% \textbf{Cryptocurrency} Nigeria's 2021 cryptocurrency ban ($n=6$) introduced complications that may explain why Nigerian participants demonstrated high cryptocurrency literacy and engagement despite regulatory prohibition. The ban forced more peer-to-peer trading and underground routes to bypass government retaliation. NP15 said: 
% \begin{quote}
%     Even now you have to be switching on VPN to use Binance...You have to do so you can change your location and that.
% \end{quote}

% Post-adoption experiences sculpt how Nigerians, Zimbabweans, and Tanzanians perceive and utilize the payment systems at their disposal. Persistent failures in financial costs, operational reliability, security, and institutional accountability manifested differently across digital currency types and national contexts. Users demonstrated remarkable financial literacy managing several systems at a time, with strategies shaped by regulatory environments, institutional trust levels, and system deficiencies.

\subsection{Takeaway messages}
\label{sec:takeaways}
Our study surfaced several themes about why people use and select between digital payment systems. 
% Three main takeaways emerge.

\paragraph{(1) While reliability and convenience contribute strongly to the continued use of payment systems, almost all of our participants experienced serious usability obstacles.}
We expected to hear about the importance of reliable service, low fees, and convenience, including the demand for physical branches and human agents. However, users were very open about the shortcomings of digital payment systems on a pure usability level, including long wait times, opaque and unpredictable fees, and lack of reliable recourse in case of mistakes or malicious activity. These shortcomings were often associated with limited infrastructure, either technological (e.g. network connectivity) or human (e.g., physical branches, human agents).

\paragraph{(2) Complex and sometimes conflicting trust relationships with the government strongly impact digital payment adoption and continued use.}  
A more surprising finding concerned the ways people think about government involvement in digital payment systems. Our results showed that some users view the endorsement of a government entity positively, helping to build trust in the legitimacy of a product. 
In particular, government endorsement gives some people confidence that they are less likely to be defrauded.
% Moreover, we found that users are positively influenced to adopt new currencies by network effects, social trust (particularly from people they know), and financial incentives. 
At the same time, many users harbor deep-rooted concerns regarding the integrity, quality of service, and/or infrastructure that governments can provide. 
Specific concerns appear to be country-dependent, as explained in \Cref{sec:country-differences}, and in some cases, we observed the same users expressing both trust and distrust in the government, for different reasons.
Such complex trust profiles can impact both the likelihood of user adoption, and also their modes of continued use of digital payment systems (e.g. not choosing to use a particular service for a particular transaction due to surveillance concerns).

%\paragraph{(3) Today, a fragmented digital payment ecosystem allows users to navigate varying, complex personal priorities and trust preferences.}  
\paragraph{(3) Today, a fragmented digital payment ecosystem leads users to navigate varying, complex personal priorities and trust preferences.}  
Many respondents to our intake survey reported having accounts with various digital payment systems at once---a phenomenon also identified in prior work \cite{kotut2024things}. 
With these accounts, users make complex and nuanced decisions about which payment system to use for different purposes, and when. Deciding factors can include accessibility (whether a particular platform is accepted by a particular vendor), fees, financial volatility, system robustness, and perceptions of data protection and recourse. 
These complex decisions are enabled today by a patchwork of digital payment systems, both private and public, each with different properties and infrastructure. 
On a more positive note, competition between providers helps users by giving them greater choice and flexibility.
% Our findings suggest that conversations around financial inclusion may benefit from additional nuance: simply having an account on any of these digital payment platforms does not necessarily allow users to participate in many kinds of financial transactions they may require in their daily lives. 

\subsection{Recommendations}
\label{sec:recommendations}
\ifdoubleblind
\else
From the above findings, we derive the following recommendations. 
\fi

\paragraph{(1) Governments should require payment providers to more clearly communicate transaction and maintenance fees to users.} 
Our findings indicate that overcharging by agents and payment providers is common, particularly in Nigeria and Zimbabwe. Similar findings have been previously observed in \cite{raghunath2024beyond} and \cite{transparency-Ng}. In the former study, the authors found agents in Tanzania sometimes imposed unofficial charges for aiding lower-literacy customers in executing transactions, while the latter study found information digital payment platforms  provided inconsistent information on fee structures.  

The recurring challenge of fee opacity highlights an urgent need for improved policies on disclosure requirements. Zimbabwe requires that transaction fees be approved by the Reserve Bank of Zimbabwe \cite{SI2020080Banking}, but there are no clear regulations on transaction fee disclosure. While Tanzania \cite{tanzania-emoney, bot-fin-protection, tanzania-microfinance} and Nigeria \cite{momo-reg-ng} have regulations stipulating transaction fee limits and disclosure requirements for digital payment providers, only Tanzania comprehensively outlines where, when and how such disclosures should happen. Specifically, Tanzania mandates that finance providers communicate transaction costs through electronic media and physical displays at providers’ offices and agents’ outlets in addition to providing an itemized breakdown of all costs before and after a customer confirms a digital transaction \cite{tanzania-emoney, bot-fin-protection, tanzania-microfinance}. Any changes to costs must be approved by BoT and communicated to customers before being operationalization \cite{tanzania-microfinance}. Enforcement of these regulations resulted in the lowering of mobile money microcredit interest rates from 10\% to 4\% \cite{credit-reg-Tz}. We advocate for widespread adoption of this model as improved transparency and accountability have  positive impacts on reducing fees, which disproportionately affect lower-income populations. 

\paragraph{(2) Governments should take an active role in supporting interoperability between platforms.} 
Several studies and reports have outlined the benefits to consumers of payment system interoperability. These include lower transaction costs \cite{brunnermeier2023mobile, ancalle2024impact}, which can lead to increased uptake of digital payments, along with users’ deeper engagement with a wider range of services \cite{interop_ipa}. Our findings suggest that supporting interoperability can be a competitive advantage for providers, as it influences users’ decisions to continue adopting one payment system over others.

Despite these mutual benefits, lack of trust between operators, a fear of losing market dominance, currency fluctuations and incompatible digital systems can hamper interoperability efforts \cite{kablanck2015achieving, domingo2023interoperability}. Tanzania and Ghana showed that these challenges can be overcome through more involvement of the government and other parties. In Tanzania, interoperability talks had stalled despite demand until the International Finance Corporation (IFC) facilitated negotiation and signing of bilateral agreements  \cite{kablanck2015achieving}. 
In Ghana, the Ghana Interbank Payment and Settlement System (GhIPSS) launched mobile money interoperability in 2018 after observing demand for third-party products like MOVE, enabling transfers between providers  \cite{CGAP-interop}. Similar interventions by governments across low- and middle-income countries could help ensure interoperability is implemented within countries and across borders. Such interventions can help manage a broader trend towards the fragmentation of global payment systems \cite{fragmented-payments}.

%Key findings reported in paper: (1) 
% should actively involve partnerships with infrastructure providers and social influence channels.  
% have important implications for the rollout of future digital payment systems, and perhaps digital public infrastructure more broadly. 
% These findings 
% We 
%suggest two requirements for the successful deployment and rollout of future digital payment systems, and possibly digital public infrastructure more broadly. The first is well-designed socialization campaigns that couple social influence (not just advertisement) with financial incentives. 
% Our study suggests that 
%The second requirement is structural: new rollouts need good infrastructure to survive. In cases where governments may not be equipped to provide such infrastructure, strong partnerships with trusted providers (e.g., telecommunication service providers) can ensure a high quality of service needed to attract and retain users. 
%For example, this model appears to be working well in Zimbabwe, where ZiG integration with popular MoMo providers (e.g., EcoCash) have boosted adoption.  
% \gf{Takudzwa, Prof. Rusero, is this true?}[TK: I would say yes because most people use ZIG through ecocash]

\paragraph{(3) HCI researchers should explore and evaluate interface design strategies and agent incentive mechanisms that prevent low digital literacy users from being exploited by intermediaries. }
Our findings suggest that some users rely consistently on intermediaries to transact. While our study did not explicitly surface why people use intermediaries, prior studies have highlighted low digital literacy and convenience as contributing factors \cite{kotut2024things}. 
However, such intermediaries pose known privacy and security risks \cite{jain2021protecting, sowon2024role}. 

There are guidelines for interfaces to support low-literacy application users, like enabling multimodal inputs such as voice/text and use of pictography and icons instead of text \cite{srivastava2021actionable, 10.1145/3578837.3578842}. 
Researchers have also explored interfaces that allow low-literacy users to get real-time remote support from their higher-literacy contacts to complete digital tasks %like making a call or saving a new contact 
\cite{suhrid2015}. However, 
% live usability evaluations of these designs have been limited; 
we do not currently understand whether these interfaces and interaction patterns increase users' capability and confidence, and empower them to independently use digital applications in practice. This is a critical gap, particularly with the increasing use and incentivization of digital payments over cash \cite{klapper2025global}, driven in part by cashless policies, e.g., in Nigeria and Zimbabwe. 
HCI researchers can play an important role in both evaluating existing frameworks longitudinally and designing improved policies and frameworks in case existing guidelines are not as effective as previously thought.

% Similarly, while there is increasing literature on user-agent relationships (e.g., \cite{sowon2024role}), few studies explore the dynamics from agents' perspectives. To mitigate user exploitation, more studies are needed on agents' motivations, incentives, and legal obligations. 
% and how these and other factors influence their interactions with users.   

\ifdoubleblind
\else
\paragraph{(4) Researchers should study the design of influence campaigns for digital public infrastructure.} 
% While our first recommendation is tailored primarily toward governments and policymakers, our second recommendation is for the research community: 
It is poorly understood how socialization and rollout campaigns impact the adoption and eventual success of digital public infrastructure. 
eNaira, for example, struggled in part due to low public awareness \cite{bosua2024public}. 
% ZiG is a really interesting case study, since it started as a CBDC, before replacing the Zimbabwean dollar as a ``fiat'' currency, but remains mostly used as a digital currency today.
Even much better-resourced countries like China have struggled  with CBDC adoption, despite greater publicity and incentives; e.g., the e-CNY had low initial adoption \cite{ecny-low-adoption} but accelerated significantly in recent years \cite{ac-alisha}.
However, we lack systematic research on how to design effective influence campaigns for public infrastructure. 
% How should governments best combine organic advertisement (e.g. referrals) with planted promotions (e.g., social media, conventional advertisement) with financial incentives (e.g., discounts, limited choice) to encourage adoption?
% Finally, are such drivers of adoption and use good for society? 
We propose that the research community study the effects of prior and ongoing rollout campaigns (e.g., comparing eNaira with ZiG with e-CNY \cite{saliya2025china}) to understand how various approaches result in (a) the spread of awareness, (b) the spread of positive sentiment, and (c) the spread of adoption.  
This could be done via a combination of  qualitative and quantitative studies.
Based on these results, the research community could propose best practices for designing rollout campaigns for public infrastructure; 
% Such proposals may 
this could build on a long literature on understanding and encouraging technology adoption (e.g., \cite{Venkatesh:MISQ03}), while accounting for the specific properties of digital payment systems. 
\fi

% that make them appealing (or not) for various use cases.
%\nicolasc{so the main thing about zig though is that in my understanding, it started as a cbdc, then they killed the fiat currency (zimbabwean dollar?) and eventually emitted "physical zig" (which they now call zig, the digital cbdc has now officially a different name i forgot but that no one seems to use) but de facto, a cbdc became the only legal tender (well, other than usd)... so that's a really strange situation, and i think it really impacts the results we got. I feel like we didn't really integrate this in the discussion. @TK and Prof. Rusero, did I get this right?}
%\gf{TK can you comment pls?}

%\tk{This is my understanding and maybe i stand to be corrected : The ZIG started off as a CBDC and the original plan was for it to be solely for investment purposes, mimicking the Mosi a Tunya Gold Coin that was introduced in 2022, but a digital one. They then introduced the ZIG but it was now used for retail purposes and later nationalized to replace the Bond Notes that were no longer functional in the economy due to inflation, thats when the ZIG cash also joined into the busket of currencies. So yah, i would say the CBDC was made the defacto currency even though up to now the USD is the dominant one}

% \paragraph{(3) New measures for financial inclusion?}
% \gf{Not sure we need this, but we could propose to measure what proportion of transactions people might want to execute in their daily lives they are able to execute (have an account that would allow them to do so). }

%% file: Sections/06-Conclusion.tex
% Conclusion
\section{Conclusion}
\label{sec:conclusion}

This work presents a qualitative examination of users' sentiments towards various digital payment systems in Tanzania, Nigeria, and Zimbabwe. 
Our findings reveal that in all three countries, users conduct a daily, complex negotiation between sometimes-conflicting priorities, including cost, convenience, security, privacy, and trust, to decide which payment systems to use. 
These internal negotiations are impacted in part by a lack of transparency and usability on the part of payment service providers. 
Our results add nuance to existing literature on the adoption and usage of digital payment systems, shedding light on \emph{how} people decide to favor one payment system over another in day-to-day interactions, beyond simple binary choices to adopt or abandon a system. 
Overall, our work suggests that given the complex landscape of digital payment systems and a growing trend of fragmentation, it is increasingly important to manage the linkages and interoperability between digital payment systems; particularly in low-income countries, partitioning money across siloed payment services poses financial and practical barriers to user financial autonomy.

%% file: Sections/07-Acknowledgments.tex
\begin{acks}
\ifdoubleblind
This paper was polished for grammar using  Grammarly, ChatGPT,
Sider AI, and Claude.

\else
This work was supported in part by the Gates Foundation, as well as the industry sponsors for the Initiative for CryptoCurrencies and Contracts (IC3) and the Carnegie Mellon University Secure Blockchain Initiative (SBI). The views and opinions expressed in this study are those of the authors and do not necessarily reflect the views or positions of the sponsors. This work was also sponsored in part by NSF grant CNS2325477. We also acknowledge the team from Telone Centre for Learning, particularly Mr. Tawanda Mapuranga for helping with the ethics approval process, participant recruitment, and allowing us to use their facilities for interviews in Zimbabwe.

\fi
\end{acks}

%% file: Appendices/00-ethics.tex
\section{Ethical Considerations}
\label{sec:ethics}

\subsection*{Impact on Participants}
This study asked participants to reflect on their own use of digital payment systems. 
Payments are of course a private and potentially sensitive topic for some users. 
For example, some users reflected on negative opinions of various service providers, including the government. 
We even had a few instances of participants sharing stories of using payment systems for activities of dubious legality. 
As such opinions could be harmful if linked back to participants, we  committed to maintaining the anonymity of all participants, and have decoupled user names from all recordings and transcriptions. 
We also communicated this to participants before the interview, to assure them that their responses would be kept private and anonymous.

\subsection*{Impact on Researchers}
This study required some of the authors to travel to perform local field work to gather interview responses. Due to the current worldwide potential for geopolitical instability, the team ensured that all authors could be repatriated immediately to their main institution if need be, and interviews were scheduled as compactly as possible to reduce time spent abroad (we did not shorten the interviews themselves, but tried to complete the 30 interviews per country in as compact of a time frame as possible). 
% This was som
%This turned out not to be necessary, even though we deliberately elected to minimize the amount of time devoted to field work, to minimize the risk for potential complications. This explains why we limited ourselves to 30~interviews per country. \gf{I'm worried this will raise questions... eg why 30. Maybe we just remove the last two sentences?}\nc{yes good point, if it gets in we can always put it back in}

\subsection*{Interviews}
\subsubsection*{Consent}
All participants provided informed verbal consent prior to participation, with consent procedures conducted in participants' preferred languages where necessary. Participation was entirely voluntary, and participants were informed of their right to withdraw from the study at any time without penalty or negative consequences.

\subsubsection*{Compensation}
 Participants received compensation of 10 USD, paid via mobile money, cash, or bank transfer depending on national norms and participant preferences. 
 We also reimbursed transportation costs up to 10 USD per participant. The compensation amount was determined to be appropriate and non-coercive by our Institutional Review Board (IRB) and each of the local ethics committees, providing fair recognition for participants' time while avoiding undue inducement.

\subsection*{Ethics Approvals}
This study was approved by our institutional review board (IRB), as well as the following local ethics boards.

\paragraph{Nigeria}
This study received approval from the 
\ifdoubleblind
[Redacted for double blind review] Ethics Committee, 
\else
University of Lagos Research Ethics Committee (UNILAGREC),
\fi
which is an ethics board housed at 
\ifdoubleblind
a local institution, 
\else
the University of Lagos
\fi
but  authorized to provide national approval for research studies  in Nigeria.
% University of Lagos Research Ethics Committee (UNILAGREC).
%This study was conducted in accordance with ethical research standards and received approval from the University of Lagos Research Ethics Committee (UNILAGREC) and our local %Carnegie Mellon University's 
%IRB. 
% UNILAGREC is an ethics committee that can provide ethics approval for any study in Nigeria; it does not require that the study be conducted at the University of Lagos. 
%\gf{Isabel, pls check}

\paragraph{Zimbabwe}
In Zimbabwe, we obtained ethical approval from the Research Council of Zimbabwe (RCZ), as mandated by the Research Act [Chapter 10:22] for all research conducted within the country \cite{research_act}. A detailed application to the RCZ was submitted, including the study protocol, informed consent forms, and data collection materials, thereby ensuring compliance with national ethical standards. The application process involved a review by the Medical Research Council of Zimbabwe (MRCZ), which assessed whether the study was medically related before final consideration by the RCZ, which evaluated the study's ethical implications, participant safety, and the mechanisms researchers had put in place to ensure participants' data protection.

\paragraph{Tanzania}
In Tanzania, the study received ethical approval from %the Carnegie Mellon University Institutional Review Board (IRB) 
our Institutional Review Board
and was registered with the Tanzania Commission for Science and Technology (COSTECH). In line with national research policy, a multitier approval process was followed. This included obtaining national clearance through COSTECH and additional site-specific permits from the President’s Office Regional Administration and Local Government (PO-RALG), which authorized research activities in Dar es Salaam, Dodoma, and Morogoro. Local government cooperation was also facilitated through letters issued to regional and district offices.

% To ensure participant confidentiality, all interview data were anonymized using coded identifiers. Personal identification information was excluded from the transcripts and recordings, which were securely stored in encrypted formats accessible only to the research team. The participants were compensated by 24,000 Tanzanian shillings (approximately 10 USD), along with reimbursement for transportation expenses, where applicable. Compensation was reviewed to ensure that it was fair, culturally appropriate, and noncoercive.

\subsection*{Data management}
To protect participant privacy and confidentiality, all interview data were anonymized using participant codes, with identifying information removed from transcripts and recordings. Audio recordings and transcripts were stored securely with encryption %during storage and transmission, with
and access limited to authorized research team members only. No individual responses are identifiable in this publication or related reports.
We have committed to deleting the associated recordings and transcriptions %
at most 3 years after the study is accepted for publication, in accordance with the researchers'   institutional IRB.

%% file: Appendices/02-interview.tex
\section{Interview Scripts}
\label{sec:interview-scripts}
\subsection{Introduction and Consent}

``Hello! Thank you for participating in this interview. Our goal is to understand how people in Africa use digital currencies like central bank digital currencies, mobile money, or cryptocurrencies and what influences their choices on payment methods. This interview will take approximately 45 minutes, with a maximum of 1 hour. Your responses will remain anonymous, meaning your responses will not have your name tied to them. We will also delete all contact information after payment. Your responses will remain confidential and will only be used for research purposes. You can stop the interview at any time without penalty. You will receive 10 USD on completing this survey. Do you have any questions so far?''

``We would also like to record this session for transcription. Your voice recording will only be used for the purposes of this study and will not be shared with anyone outside of us researchers. We will also delete all voice recordings at the conclusion of our study, and you may ask us to stop recording at any time. Do you consent to voice recording and want to proceed?''

\subsection{Demographics}

``Before we get started, there are a few questions on the provided handout that we would like you to fill out. These demographic questions are completely optional to answer; any question you do not wish to answer can be left blank.''

\textit{[Allow participants to fill out the demographics section at their own convenience.]}

\subsection{Digital Payment Usage}

``Now, I will provide you with a list of some digital payment methods available in your region. Please check the boxes of all the digital currencies you currently use or have used. If anything is unclear or you have any questions about what is listed on the sheet, please do not hesitate to ask!''

\textit{[Allow participant to fill out the digital payment usage section at their own convenience.]}

\subsection{Per-Service Questions}

``Thank you for filling out the demographics sheet and the digital currency usage document! I will now proceed with questions regarding your exposure to each of the categories you checked on the sheet. If any questions come up while I talk, please feel free to interrupt me and ask!''

\textit{[Randomly select a category that the participant has checked. For each response, ensure you know which product in the category they are discussing.]}

\begin{enumerate}
    \item Can you tell me about who introduced you to this service and why you started using this service?
    \begin{itemize}
        \item \textit{If the user mentions hearing about it from someone, ask for more details:}
        \begin{itemize}
            \item Who are they?
            \item How close are you?
            \item Do you trust the person?
        \end{itemize}
        \item \textit{If they give a non-answer, ask how they were introduced to the service.}
    \end{itemize}
    
    \item How trustworthy do you think the issuer (or service provider for money transfer platforms)---that is, the person or organization who supplies and distributes the digital currency---of this service is?
    \begin{itemize}
        \item Do you trust them to keep your money and your personal information safe? Why?
    \end{itemize}
    
    \item What kind of transactions (investments, sending money, exchanging currencies, etc.) do you use this service for and why?
    
    \item What are the advantages of using this service? \textit{[Ask them to specify which service they are speaking about.]}
    
    \item Can you describe a situation where you chose to use a specific product from this category over another digital payment service and why? You may choose to compare within categories or between different categories.
    \begin{itemize}
        \item \textit{Skip this question if they have already answered it previously.}
        \item \textit{If they ask for clarification, explain that if they have used multiple products from the same category, they can compare those.}
    \end{itemize}
    
    \item What are the disadvantages of using this service? \textit{[Ask them to specify which service they are speaking about.]}
    \begin{itemize}
        \item Can you describe a situation where using the service was harder than expected? \textit{[Skip if they already told a story.]}
    \end{itemize}
    
    \item If you were to be scammed out of money while using this service, what kind of help would you expect to receive and how?
\end{enumerate}

\subsection{Comparison of Services}

``Between all of the categories of services you said you have used, which do you find...''

\textit{[Remind them of the categories they checked.]}

\begin{enumerate}
    \item Easiest to use?
    \begin{itemize}
        \item Why?
    \end{itemize}
    
    \item Most trustworthy?
    \begin{itemize}
        \item Why?
    \end{itemize}
    
    \item Most useful for your daily life?
    \begin{itemize}
        \item Why?
    \end{itemize}
\end{enumerate}

\subsection{Hypothetical Scenarios}

``Now I'd like to explore some hypothetical scenarios with you. These aren't real situations, but imaginary ones that help us understand how you might make decisions about digital payment services. There are no right or wrong answers---we're simply interested in your thoughts and reasoning.''

\subsubsection{Scenario 1: Government vs. Peer-Recommended Platform}

``Imagine two new digital payment systems that have just launched. One is issued by your government, but you have not heard anything from your peers about it. The other is issued by an unknown party, but your friend tells you it is a good product. Both offer similar features: fast transactions and low fees, and both are accepted by many merchants.''

\begin{itemize}
    \item Which of these would you be more likely to try, if any? Why?
    \item What else would convince you to try out the platform you selected?
\end{itemize}

\subsubsection{Scenario 2: Government vs. Tech Company Platform}

``Imagine two new digital payment systems that have just launched. One is issued by your government, and the other by a well-known tech company that you already use for things like messaging or payments (such as MTN or Vodacom). Both offer similar features: fast transactions and low fees, and both are accepted by many merchants.''

\begin{itemize}
    \item Which of these would you be more likely to try, if any? Why?
    \item What else would convince you to try out the platform you selected?
\end{itemize}

\subsubsection{Scenario 3: Failed Transaction Without Recovery}

``Imagine you are using a digital payment system you have used before. You try to pay for something online, but the transaction fails, and you lose your money. You try to contact someone for help, but you're not able to recover the funds.''

\begin{itemize}
    \item Has this ever happened to you? What happened?
    \item Would you ever use this service again?
    \item If this happened to you, why did you make your decision?
    \item What would convince you to use it again?
\end{itemize}

\subsubsection{Scenario 4: Failed Transaction With Recovery}

``Imagine you are using a digital payment system you have used before. You try to pay for something online, but the transaction fails, and you lose your money. However, after contacting an authorized party, you are able to recover your money.''

\begin{itemize}
    \item Would you use this service again? Why?
    \item Has something like this happened to you or someone you know?
\end{itemize}

\subsection{Closing}

``That concludes our interview. Thank you so much for sharing your experiences and insights with us today. Your responses will be invaluable for our research on digital currency usage in Africa. As mentioned at the beginning, all your information will remain anonymous and confidential. If you have any questions about the study, if you would like a copy of your responses, or would like to know about our findings once the research is complete, please don't hesitate to contact us. Thank you again for your time and participation!''

\label{app:interview}

%% file: Appendices/03-codebook.tex
\section{Codebook}
\label{app:codebook}

\ifdoubleblind
Our codebook can be accessed at  
\url{https://osf.io/ue73z/files/cnx98?view_only=7d4758f9c43d41ddbe2ac55b0e9b1ce6}.
\else
Our codebook can be accessed at  
\url{https://osf.io/ue73z/files/cnx98?view_only=7d4758f9c43d41ddbe2ac55b0e9b1ce6}.
\fi

%% file: Bibliography/Bibliography.bib
@inproceedings{kotut2024things,
  title={" Things on the Ground are Different": Utility, Survival and Ethics in Multi-Device Ownership and Smartphone Sharing Contexts},
  author={Kotut, Lindah and Alikhan, Hummd},
  booktitle={Conference on Human Factors in Computing Systems (CHI)},
  pages={1--14},
  year={2024}
}

@article{hinman2010opportunities,
  title={Opportunities and challenges for mobile-based financial services in rural Uganda},
  author={Hinman, Rachel and Matovu, Julus},
  journal={Extended Abstracts on Human Factors in Computing Systems (CHI)},
  pages={3925--3930},
  year={2010}
}

@inproceedings{mustafa2019digital,
  title={Digital financial needs of micro-entrepreneur women in Pakistan: is mobile money the answer?},
  author={Mustafa, Maryam and Mazhar, Noor and Asghar, Ayesha and Usmani, Maryem Zafar and Razaq, Lubna and Anderson, Richard},
  booktitle={Conference on Human Factors in Computing Systems (CHI)},
  pages={1--12},
  year={2019}
}

@article{nissenbaum2004privacy,
  title={Privacy as contextual integrity},
  author={Nissenbaum, Helen},
  journal={Washington Law Review},
  volume={79},
  pages={119},
  year={2004},
  publisher={HeinOnline}
}

@article{shen2020can,
  title={'I Can't Even Buy Apples If I Don't Use Mobile Pay?' When Mobile Payments Become Infrastructural in China},
  author={Shen, Hong and Faklaris, Cori and Jin, Haojian and Dabbish, Laura and Hong, Jason I},
  journal={Proceedings of the ACM on Human-Computer Interaction},
  volume={4},
  number={CSCW2},
  pages={1--26},
  year={2020},
  publisher={ACM New York, NY, USA}
}

@article{zhou2011effect,
  title={The effect of initial trust on user adoption of mobile payment},
  author={Zhou, Tao},
  journal={Information development},
  volume={27},
  number={4},
  pages={290--300},
  year={2011},
  publisher={Sage Publications Sage UK: London, England}
}

@inproceedings{hillman2014user,
  title={User challenges and successes with mobile payment services in North America},
  author={Hillman, Serena and Neustaedter, Carman and Oduor, Erick and Pang, Carolyn},
  booktitle={Proceedings of the 16th international conference on Human-computer interaction with mobile devices \& services (MobileHCI)},
  pages={253--262},
  year={2014}
}

@inproceedings{sas2015exploring,
  title={Exploring trust in Bitcoin technology: a framework for HCI research},
  author={Sas, Corina and Khairuddin, Irni Eliana},
  booktitle={Proceedings of the Annual Meeting of the Australian Special Interest Group for Computer Human Interaction},
  pages={338--342},
  year={2015}
}

@article{fragmented-payments,
author = {Brownstein, Greg},
  title = {Global payment systems are fragmenting. Here's what the G20 can do},
  journal = {Atlantic Council},
  url = "https://www.atlanticcouncil.org/in-depth-research-reports/issue-brief/global-payment-systems-are-fragmenting-heres-what-the-g20-can-do/",
month = {4},
year = {2026},
  note = "[Online; accessed 2026-04-10]"
}

@MastersThesis{munthali2024impact,
  title="Impact of Digital Transformation in the Remittances Sector in {Zimbabwe}",
  author={Munthali, Ian},
  year={2024},
  school={Africa University},
  url = {http://41.174.125.165:4024/jspui/bitstream/123456789/4499/1/Munthali%2C%20Ian.%202024.%20Impact%20of%20Digital%20Transformation%20in%20the%20Remittances%20Sector%20in%20Zimbabwe.pdf}
}

@article{avom2023financial,
  title={Do financial innovations improve financial inclusion? Evidence from mobile money adoption in Africa},
  author="Avom, D{\'e}sir{\'e} and Bangak{\'e}, Chrysost and Ndoya, Hermann",
  journal={Technological Forecasting and Social Change},
  volume={190},
  pages={122451},
  year={2023},
  publisher={Elsevier},
  doi = {10.1016/j.techfore.2023.122451}
}

@article{abdurrahaman2024revisiting,
  title="Revisiting the Factors Affecting Cryptocurrency Adoption: Evidence from {Nigeria}",
  author={Abdurrahaman, Daha Tijjani and Ayetigbo, Olumide Abiodun and Okunlola, Olalekan Charles and Adegbola, Abimbola Eunice},
  journal={Journal of African Business},
  pages={1--28},
  year={2024},
  volume = 27, 
  number =1, 
  publisher={Taylor \& Francis},
  doi = {10.1080/15228916.2024.2432695}
}

@article{opay,
author = {Ojeniyi, Sekinat Motunrayo},
  title = "The app that revolutionized money transfers in {Nigeria}",
  journal = {Rest of World},
month = {10},
year = {2023},
  note = "[Online; accessed 2026-02-05]",
  url = {https://restofworld.org/2023/opay-app-nigeria-money-transfers/}
}

@article{PalmPay,
author = {Bright, Jake },
  title = {{PalmPay launches in Nigeria on \$40M round led by China’s Transsion}},
  journal = {TechCrunch},
month = {11},
year = {2019},
url = {https://techcrunch.com/2019/11/12/palmpay-launches-in-nigeria-on-40m-round-led-by-chinas-transsion/},
  note = "[Online; accessed 2026-02-05]"
}

@article{yaqub2013cashless,
  title="The cashless policy in {Nigeria}: prospects and challenges",
  author={Yaqub, J.O. and Bello, H.T. and Adenuga, I.A. and Ogundeji, M.O.},
  journal={International Journal of Humanities and Social Science},
  volume={3},
  number={3},
  pages={200--212},
  year={2013},
  month = feb
}

@article{luhanga2023user,
  title="User Experiences with Third-Party {SIM} Cards and {ID} Registration in Kenya and Tanzania",
  author={Luhanga, Edith and Sowon, Karen and Cranor, Lorrie Faith and Fanti, Giulia and Tucker, Conrad and Gueye, Assane},
  journal={Conference on Human Factors in Computing Systems (CHI)},
  year={2026}
}

@article{ac-alisha,
  author = {Chhangani, Alisha},
  title = "What to watch as {China} prepares its digital yuan for prime time",
  journal = {Atlantic Council},
  month = {1},
  year = {2026},
  url = {https://www.atlanticcouncil.org/blogs/econographics/what-to-watch-as-china-prepares-its-digital-yuan-for-prime-time/},
  note = "[Online; accessed 2026-02-05]"
}

@misc{ecny-low-adoption,
  title = "China to Pay Interest on Digital Yuan in Bid to Boost Adoption",
  howpublished = {Bloomberg News},
  month = {12},
  year = {2025},
  note = "[Online; accessed 2026-02-05]"
}

@article{saliya2025china,
  title="China's {e-CNY} and the Future of Money: A Rewiring Global Finance, Technology, Policy, and Geopolitics",
  author={Saliya, Candauda Arachchige},
  year={2025},
  month = jul, 
  howpublished = {SSRN}, 
  url = {https://ssrn.com/abstract=5350710},
  doi = {10.2139/ssrn.5350710}
}

@article{akinyemi2020determinants,
  title="Determinants of mobile money technology adoption in rural areas of {Africa}",
  author="Akinyemi, {Babatope E.} and Mushunje, Abbyssinia",
  journal={Cogent Social Sciences},
  volume={6},
  number={1},
  pages={1815963},
  year={2020},
  publisher={Taylor \& Francis},
  doi = {10.1080/23311886.2020.1815963}
}

@article{mothobi2017infrastructure,
  title="Infrastructure deficiencies and adoption of mobile money in {Sub-Saharan Africa}",
  author="Mothobi, Onkokame and Grzybowski, Lukasz",
  journal={Information Economics and Policy},
  volume={40},
  pages={71--79},
  year={2017},
  publisher={Elsevier},
  doi = {10.1016/j.infoecopol.2017.05.003}
}

@article{ngo2023governance,
  title="Governance and monetary policy impacts on public acceptance of {CBDC} adoption",
  author={Ngo, Vu Minh and Van Nguyen, Phuc and Nguyen, Huan Huu and Tram, Huong Xuan Thi and Hoang, Long Cuu},
  journal={Research in International Business and Finance},
  volume={64},
  pages={101865},
  year={2023},
  publisher={Elsevier},
  doi = {10.1016/j.ribaf.2022.101865}
}

@misc{Egobiambu2025DigitalPaymentsNigeria,
  author       = {Egobiambu, Emmanuel},
  title        = "How Digital Payments Are Transforming Businesses in {Nigeria}",
  howpublished = {\url{https://www.channelstv.com/2025/12/10/how-digital-payments-are-transforming-businesses-in-nigeria/}},
  journal      = {Channels Television},
  year         = {2025},
  month        = dec,
  note         = {[Online; accessed 06-Feb-2026]}
}

@article{mutiso2020assessment,
  title="An assessment of the adoption of cryptocurrency as a mode of payment by {SMEs} in {Kiambu County, Kenya}",
  author={Mutiso, Agnes and Maguru, Brian},
  journal={International Journal of Social Science and Economic Research},
  volume=5,
  number=7,
  pages={2000--2013},
  year=2020,
  doi = {10.46609/IJSSER.2020.v05i07.023}
}

@article{bijlsma2021triggers,
  title={What triggers consumer adoption of central bank digital currency?},
  author={Bijlsma, Michiel and van der Cruijsen, Carin and Jonker, Nicole and Reijerink, Jelmer},
  journal={Journal of Financial Services Research},
  volume={65},
  number={1},
  pages={1--40},
  year={2024},
  publisher={Springer}
}

@article{mazambani2020predicting,
  title="Predicting {FinTech} innovation adoption in {South Africa}: the case of cryptocurrency",
  author={Mazambani, Last and Mutambara, Emmanuel},
  journal={African Journal of Economic and Management Studies},
  volume={11},
  number={1},
  pages={30--50},
  year={2020},
  doi = {10.1108/AJEMS-04-2019-0152},
  publisher={Emerald Publishing Limited}
}

@misc{bog-cbdc,
author = {Bank of Ghana},
title = {{Press Release – BOG Partners with Giesecke+Devrient to Pilot Digital Currency in Ghana}},
url = "https://www.bog.gov.gh/news/press-release-bank-of-ghana-partners-with-gieseckedevrient-to-pilot-first-general-purpose-central-bank-digital-currency-in-africa/",
month = {},
year = {},
note = "[Online; accessed 2026-02-05]"
}

@misc{vanteutem,
author = {van Teutem, Simon},
  title = {There are now more than half a billion mobile money accounts in the world, mostly in Africa -- here's why this matters},
  journal={Our World in Data},
  month = jul,
  year = 2025,
  url = {https://ourworldindata.org/mobile-money-why-it-matters},
  note = "[Online; accessed 2026-02-05]"
}

@article{e-creator,
    author = {Mutandiro, Kimberly},
    title = "A {Ponzi} scheme targets desperate workers amid {Zimbabwe}'s employment crisis",
    journal = {Rest of World},
    year = {2023},
    month = jul, 
    url ={https://restofworld.org/2023/e-creator-ponzi-scheme-zimbabwe-workers/}
}

@article{de2023analysis,
  title="An analysis of the public consequences of cybersecurity incidents in {Brazil}",
  author={de Lemos, Vit{\'o}ria and Ignaczak, Luciano},
  journal={Social Network Analysis and Mining},
  volume={13},
  number={1},
  pages={106},
  year={2023},
  publisher={Springer},
  doi = {10.1007/s13278-023-01113-9}
}

@InProceedings{rebensky2021user,
  author="Rebensky, Summer
  and Carroll, Meredith
  and Nakushian, Andrew
  and Chaparro, Maria
  and Prior, Tricia",
  title="Understanding the Last Line of Defense: Human Response to Cybersecurity Events",
  booktitle="Proc. HCI for Cybersecurity, Privacy and Trust: Third International Conference (HCI-CPT 2021)",
  year="2021",
  pages = "353-366",
  doi = {10.1007/978-3-030-77392-2_23}
}

@inproceedings{bosua2024public,
  title={Public Perception and Adoption Approaches for Digital Currencies: Analysing Influencing Factors},
  author={Bosua, Amarachukwu Patience and Biswas, Mriganka},
  booktitle={2024 29th International Conference on Automation and Computing (ICAC)},
  pages={1--6},
  year={2024},
  organization={IEEE},
  doi = {10.1109/icac61394.2024.10718745}
}

@misc{age-worldbank,
author = {{World Bank Gender Data Portal}},
  title = {{Population (age group as \% of total population)}},
  url = "https://genderdata.worldbank.org/en/indicator/sp-pop-zs?age=15-64",
  note = "[Online; accessed 2026-02-04]"
}

@misc{tz-internet,
author = {Sehloho, Matshepo},
  title = "{Tanzania} cuts {Internet} amid election day protests",
  journal={Connecting Africa},
  month = oct,
  year = {2025},
  url = {https://www.connectingafrica.com/connectivity/tanzania-cuts-internet-amid-election-day-protests},
  note = "[Online; accessed 2026-02-04]"
}

@misc{tzlgbtq,
  author = {Equaldex},
  title = {{LGBT rights in Tanzania}},
  url = {https://www.equaldex.com/region/tanzania\#non-binary-gender-recognition},
  note = "[Online; accessed 2026-02-01]"
}

@misc{zimlgbtq,
  author = {Equaldex},
  title = "{LGBT rights in Zimbabwe}",
  url = {https://www.equaldex.com/region/zimbabwe},
  note = "[Online; accessed 2026-02-01]"
}

@misc{nglgbtq,
  author = {Equaldex},
  title = {{LGBT Rights in Lagos}},
  url = {https://www.equaldex.com/region/lagos},
  note = "[Online; accessed 2026-02-01]"
}

@techreport{eib2024fintech,
  author      = {{European Investment Bank}},
  title       = "{EIB} Finance in {Africa} 2024: Fintech Transforms {African} Financial Services, but High Funding Costs Hinder Climate and Digital Transitions",
  year        = {2024},
  url         = {https://www.eib.org/en/press/all/2024-435-eib-finance-in-africa-2024-fintech-transforms-african-financial-services-but-high-funding-costs-hinder-climate-and-digital-transitions},
  note        = {[Online; accessed 06-February-2026]}
}

@article{bitrus2021mobilepayment,
  title   = "Factors Influencing the Success of Mobile Payment in Developing Countries: A Comparative Analysis of {Nigeria and Kenya} Mobile Payment Users",
  author  = {Bitrus, Stephen-Aruwan and Lee, Chol-Ho and Rho, Jae-Jeung and Erdenebold, Tumennast},
  journal = {Asia-Pacific Journal of Business},
  volume  = {12},
  number  = {3},
  year    = {2021},
  pages   = "1--36", 
  doi     = {10.32599/apjb.12.3.202109.1}
}

@article{Osuma2025,
  author    = {Osuma, Godswill},
  title     = {The Impact of Financial Inclusion on Poverty Reduction and Economic Growth in Sub-Saharan Africa: A Comparative Study of Digital Financial Services},
  journal   = {Social Sciences \& Humanities Open},
  year      = {2025},
  volume    = {Volume 11},
  number    = {ISSN 2590-2911},
  pages     = {},
  doi       = {10.1016/j.ssaho.2024.101263},
  month     = {},
  note      = {}
}

@article{harris2013,
  title="Privacy and Security Concerns Associated with Mobile Money Applications in {Africa}", 
  author = "Andrew Harris and Seymour Goodman and Patrick Traynor", 
  journal= " Wash. J. L. Tech. \& Arts",
  pages= "245--264",
  volume =8, 
  number =3,
  year=2013
}

@misc{bot-fin-protection,
  author = {Bank of Tanzania},
  title     = {Financial Consumer Protection Regulations},
  year      = {2019},
  url       = {https://www.bot.go.tz/Publications/Acts,%20Regulations,%20Circulars,%20Guidelines/Regulations/en/2020031802343226.pdf},
  urldate   = {2026-03-02},
  note = "Tanzania financial consumer protection law"
}

@misc{tanzania-emoney,
   author = {Bank of Tanzania},
  title     = {The Electronic Money Regulations},
  year      = {2015},
  url       = {https://www.bot.go.tz/Publications/NPS/GN-THE%20ELECTRONIC%20MONEY%20REGULATIONS%202015.pdf},
  urldate   = {2026-03-02},
  note = "Tanzania electronic money law"
}

@misc{tanzania-microfinance,
  author = {Bank of Tanzania},
  title     = {Guidelines on Fees and Charges for Microfinance Service Providers},
  year      = {2024},
  url       = {https://www.bot.go.tz/Publications/Acts,%20Regulations,%20Circulars,%20Guidelines/Guidelines/en/2024071716375097.pdf},
  urldate   = {2026-03-02}
}

@techreport{credit-reg-Tz,
    author = {Thiemele-Kadjo, Ruth and Priollaud, Simon and Kleiman, David and Sanchez, Alexandra and Dadi , Bernard and Mohamma, Ghiyazuddin and Parvez, Jaheed},
    title = {Digital Credit Regulation in Tanzania},
    institution = {Alliance for Financial Inclusion},
    year = {2020},
    url={https://www.afi-global.org/wp-content/uploads/2024/10/AFI_DFS_Tanzania_CS_AW2_05.10.20_digital.pdf},
    urldate   = {2026-03-02}
}

@techreport{transparency-Ng,
    author = {Blackmon, William and Mwesigwa, Brian},
    title = {Measuring Fees and Transparency in Nigeria’s Digital Financial Services},
    institution = {Innovations for Poverty Action},
    year = {2022},
    url={https://poverty-action.org/sites/default/files/2023-05/Report_Measuring-Fees-Transparency-Nigeria-Digital-Finance_2022.02.16_Project-Page.pdf},
    urldate   = {2026-03-02}
}

@misc{momo-reg-ng,
  author = {Central Bank of Nigeria},
  title     = {Regulatory Framework for Mobile Money Services in Nigeria},
  year      = {2021},
  url       = {https://www.cbn.gov.ng/Out/2021/CCD/Framework%20and%20Guidelines%20on%20Mobile%20Money%20Services%20in%20Nigeria%20-%20July%202021.pdf},
  urldate   = {2026-03-02}
}

@article{raghunath2024beyond,
  title={Beyond digital financial services: Exploring mobile money agents in Tanzania as general ICT intermediaries},
  author={Raghunath, Ananditha and Ndubuisi-Obi Jr, Innocent and Mpogole, Hosea and Anderson, Richard},
  journal={ACM Journal on Computing and Sustainable Societies},
  volume={2},
  number={1},
  pages={1--26},
  year={2024},
  publisher={ACM New York, NY}
}

@inproceedings{jain2021protecting,
  title={“Who is protecting us? No one!” Vulnerabilities Experienced by Low-Income Indian Merchants Using Digital Payments},
  author={Jain, Pranjal and Varanasi, Rama Adithya and Dell, Nicola},
  booktitle={Proceedings of the 4th ACM SIGCAS conference on computing and sustainable societies},
  pages={261--274},
  year={2021},
  url={https://doi.org/10.1145/3460112.3471961}
}

@article{srivastava2021actionable,
  title={Actionable UI design guidelines for smartphone applications inclusive of low-literate users},
  author={Srivastava, Ayushi and Kapania, Shivani and Tuli, Anupriya and Singh, Pushpendra},
  journal={Proceedings of the ACM on Human-Computer Interaction},
  volume={5},
  number={CSCW1},
  pages={1--30},
  year={2021},
  publisher={ACM New York, NY, USA}
}

@inproceedings{10.1145/3578837.3578842,
author = {Guimar\~{a}Es, Leonor and Martins, Nuno and Pereira, Leonardo and Penedos-Santiago, Eliana and Brand\~{a}O, Daniel},
title = {Interface design guidelines for low literature users: a literature review},
year = {2023},
publisher = {ACM},
address = {New York, NY, USA},
url = {https://doi.org/10.1145/3578837.3578842},
booktitle = {Proceedings of the 2022 6th International Conference on Education and E-Learning},
pages = {29–35},
numpages = {7},
series = {ICEEL '22}
}

@inproceedings{suhrid2015,
author = {Ahmed, Syed Ishtiaque and Zaber, Maruf Hasan and Morshed, Mehrab Bin and Ismail, Md. Habibullah Bin and Cosley, Dan and Jackson, Steven J.},
title = {Suhrid: A Collaborative Mobile Phone Interface for Low Literate People},
year = {2015},
booktitle={Proceedings of the 2015 Annual Symposium on Computing for Development},
publisher = {ACMy},
address = {New York, NY, USA},
url = {https://doi.org/10.1145/2830629.2830640},
pages = {95–103},
numpages = {9},
series = {DEV '15}
}

@techreport{ancalle2024impact,
  title={Impact of interoperability regulation on the use of digital payments in Peru},
  author={Ancalle, Celene and Garcia, Maria Gracia},
  year={2024},
  institution={Graduate Institute of International and Development Studies Working Paper},
  url={https://bccprogramme.org/wp-content/uploads/2024/02/HEIDWP02-2024-Ancalle-Garcia.pdf}
}

@techreport{brunnermeier2023mobile,
  title={Mobile money, interoperability, and financial inclusion},
  author={Brunnermeier, Markus K and Limodio, Nicola and Spadavecchia, Lorenzo},
  year={2023},
  institution={National Bureau of Economic Research},
  url={https://www.nber.org/system/files/working_papers/w31696/w31696.pdf}
}

@techreport{interop_ipa,
  title={How do instant interoperable payment systems transform modern economies?},
  author={Özyilmaz, Hakan},
  year={2024},
  institution={Toulouse School of Economics and Innovations for Poverty Action},
  url={https://poverty-action.org/sites/default/files/2024-11/2024-10_fit_in_initiative_how_iips_transform_modern_eco.pdf}
}

@techreport{kablanck2015achieving,
  title={Achieving interoperability in Mobile Financial Services. Tanzania Case Study},
  author = {{International Finance Corporation}},
  institution={International Finance Corporation, World Bank Group},
  address     = {Washington, DC},
  number      = {128225},
  year={2015},
  url={http://documents.worldbank.org/curated/en/740981531310065590}
}

@techreport{domingo2023interoperability,
  title={Interoperability of digital payment systems: Lessons from the East African Community},
  author={Domingo, Ennatu and Arnold, Stephanie and Apiko, Philomena},
  year={2023},
  institution={ECDPM Discussion Paper},
  url={https://atdf.trademarkafrica.com/wp-content/uploads/2024/11/Session-8-Frictionless-Frontiers-Bruce.pdf}
}

@techreport{CGAP-interop,
  title={Building Inclusive Payment Ecosystems in Tanzania and Ghana},
  author={Mattern, Max and McKay, Claudia},
  year={2018},
  institution={CGAP},
  type= {CGAP Focus Note},
  number= {110},
  url={https://hdl.handle.net/10986/30274},
}

@inproceedings{sowon2025,
	author = {Sowon, Karen and Munyendo, Collins W. and Klucinec, Lily and Maingi, Eunice and Suleh, Gerald and Cranor, Lorrie Faith and Fanti, Giulia and Tucker, Conrad and Gueye, Assane}, 
	title = "Design and Evaluation of Privacy-Preserving Protocols for Agent-Facilitated Mobile Money Services in {Kenya}",
	booktitle = "Symposium on Usable Privacy and Security (SOUPS)",
    organization={USENIX},
	pages = "391--413",
    year = {2025}
}

@misc{McKinsey2025, 
  author    = {McKinsey },
  title     = {The Future of Payments in Africa},
  year      = {2022},
  url       = {https://www.mckinsey.com/industries/financial-services/our-insights/the-future-of-payments-in-africa},
  urldate   = {2025-07-15},
}

@misc{research_act, 
  author    = "Ministry of Labour, Zimbabwe",
  title     = "Research act [Chapter 10:22]",
  year      = 2020,
  url       = {https://www.law.co.zw/download/research-act/},
  urldate   = {2025-07-16}
}

@article{soutter2019digital,
  title={Digital payments: Impact factors and mass adoption in sub-saharan Africa},
  author={Soutter, Leigh and Ferguson, Kenzie and Neubert, Michael},
  journal={Technology Innovation Management Review},
  volume={9},
  number={7},
  year={2019},
    doi = {10.22215/timreview/1254}
}

@article{makulilo2015privacy,
  title="Privacy in mobile money: Central banks in {Africa} and their regulatory limits",
  author={Makulilo, Alex B.},
  journal={International Journal of Law and Information Technology},
  volume={23},
  number={4},
  pages={372--391},
  year={2015},
  publisher={Oxford University Press},
  doi = {10.1093/ijlit/eav014}
}

@inproceedings{sowon2024role,
  title="The Role of User-Agent Interactions on Mobile Money Practices in {Kenya and Tanzania}",
  author={Sowon, Karen and Luhanga, Edith and Cranor, Lorrie Faith and Fanti, Giulia and Tucker, Conrad and Gueye, Assane},
  booktitle={Symposium on Security and Privacy (SP)},
  pages={752--769},
  year={2024},
  organization={IEEE}
}

@techreport{klapper2025global,
  title={The Global Findex Database 2025: Connectivity and Financial Inclusion in the Digital Economy},
  author={Klapper, Leora and Singer, Dorothe and Starita, Laura and Norris, Alexandra},
  year={2025},
  institution={World Bank, Washington, DC},
  doi= {10.1596/978-1-4648-2204-9}
}

@article{fabregasMobileMoneyEconomic2022,
	title = {Mobile {Money} and {Economic} {Activity}: {Evidence} from {Kenya}},
	volume = {36},
	issn = {0258-6770},
	shorttitle = {Mobile {Money} and {Economic} {Activity}},
	url = {https://pmc.ncbi.nlm.nih.gov/articles/PMC11364346/},
	doi = {10.1093/wber/lhac007},
	number = {3},
	urldate = {2026-02-03},
	journal = {The World Bank economic review},
	author = {Fabregas, Raissa and Yokossi, Tite},
	month = aug,
	year = {2022},
	pmid = {39220372},
	pmcid = {PMC11364346},
	pages = {734--756},
}

@article{omotuboraSameNairaMore2024,
	title = {Same {Naira}, {More} {Possibilities}! {Assessing} the {Legal} {Status} of the {eNaira} and {Its} {Potential} for {Privacy} and {Inclusion}},
	volume = {68},
	issn = {0021-8553, 1464-3731},
	url = {https://www.cambridge.org/core/journals/journal-of-african-law/article/same-naira-more-possibilities-assessing-the-legal-status-of-the-enaira-and-its-potential-for-privacy-and-inclusion/19189464219A6786DE28C68B5E28D67D},
	doi = {10.1017/S0021855324000044},
	language = {en},
	number = {2},
	urldate = {2026-02-04},
	journal = {Journal of African Law},
	author = {Omotubora, Adekemi},
	month = jun,
	year = {2024},
	pages = {245--262},
}

@misc{domorWhySubSaharanAfrica2025,
	title = "Why {Sub}-{Saharan} {Africa} leads in crypto adoption despite fragile markets",
	url = {https://www.globalsouthworld.com/article/why-sub-saharan-africa-leads-in-crypto-adoption-despite-fragile-markets},
	language = {en},
	urldate = {2026-02-04},
	journal = {Global South World},
	author = {Domor, Believe},
	month = aug,
	year = {2025},
}

@misc{ImpactDigitalLiteracy,
	author = "Nabil Adel",
	title = "The impact of digital literacy and technology adoption on financial inclusion in {Africa}, {Asia}, and {Latin} {America}",
	journal = {Heliyon},
	volume = 10,
	number = 24,
	pages = {e40951},
	year = {2024},
	issn = {2405-8440},
	doi = {https://doi.org/10.1016/j.heliyon.2024.e40951},
	url = {https://www.sciencedirect.com/science/article/pii/S2405844024169824}
}

@article{mndeme2025enablers,
  title   = "Enablers for Willingness to Utilize Mobile Money by Small-Scale Entrepreneurs in {Tanzania}: The Moderating Role of Payment Infrastructure",
  author  = {Mndeme, Ramadhani and Mwemezi, Justus},
  journal = {The Accountancy and Business Review},
  volume  = 17,
  number  = 1,
  year    = {2025}
}

@article{ndekwa2018adoption,
  title     = {Adoption of Mobile Money Services among University Students in Tanzania},
  author    = {Ndekwa, Beatrice and Ochumbo, Alex Juma and Ndekwa, Alberto Gabriel and John, Kalugendo Elizeus},
  journal   = {International Journal of Advanced Engineering, Management and Science},
  volume    = {4},
  number    = {3},
  pages     = "149--157", 
  year      = {2018},
  publisher = {Infogain Publication}
}

@article{lissah2024cashless,
  title     = "Cashless Payments: Perceived Challenges by Stakeholders in Tanzania Using {UTAUT2} Model",
  author    = {Lissah, Jane and Kirobo, Abdulkadir and Kaaya, Peter},
  journal   = {FUOYE Journal of Engineering and Technology},
  volume    = {9},
  number    = {1},
  pages     = {70--75},
  year      = {2024},
  doi       = {10.4314/fuoyejet.v9i1.11},
  publisher = {Faculty of Engineering, Federal University Oye-Ekiti}
}

@article{prodanRisePopularityCentral2024,
	title = {The rise in popularity of central bank digital currencies. {A} systematic review},
	volume = {10},
	issn = {2405-8440},
	url = {https://pmc.ncbi.nlm.nih.gov/articles/PMC11096978/},
	doi = {10.1016/j.heliyon.2024.e30561},
	number = {9},
	urldate = {2026-02-04},
	journal = {Heliyon},
	author = {Prodan, Silvana and Konh\"{a}usner, Peter and Dabija, Dan-Cristian and Lazaroiu, George and Marincean, Leonardo},
	month = apr,
	year = 2024,
	pmid = {38756603},
	pmcid = {PMC11096978},
	pages = {e30561},
}

@misc{UAEsDigitalDirham2025,
	author = "Prashant Jha", 
	title = {{UAE}'s {Digital} {Dirham} {CBDC} {Pilot} {Goes} {Live} -- {Global} {List} of {Launched}, {Piloting}, and {Developing} {CBDCs}},
	url = {https://www.ccn.com/news/crypto/global-list-launched-piloting-developing-cbdcs/},
	language = {en-US},
	urldate = {2026-02-04},
	journal = {CCN.com},
	month = nov,
	year = {2025},
}

@article{zhang2019security,
  title={Security and privacy on blockchain},
  author={Zhang, Rui and Xue, Rui and Liu, Ling},
  journal={ACM Computing Surveys (CSUR)},
  volume={52},
  number={3},
  pages={1--34},
  year={2019},
  publisher={ACM New York, NY, USA},
  doi = {10.1145/3316481}
}

@article{alberolaCentralBankDigitalb,
	title = {Central bank digital currencies in {Africa}},
	language = {en},
	number = {128},
	author = {Alberola, Enrique and Mattei, Ilaria},
	journal = "{BIS} Papers",
	year = 2022,
	month = nov
}

@article{alberolaCentralBankDigitalc,
	title = "Central Bank digital currencies in {Africa}: catching up",
	number = {527},
	year = 2023,
	month = feb, 
	journal = "{SUERF Policy Brief}",
	author = {Alberola, Enrique and Mattei, Ilaria}
}

@article{ricciCentralBankDigital2024a,
	title = "Central {Bank} {Digital} {Currency} and {Other} {Digital} {Payments} in {Sub}-{Saharan} {Africa}",
	volume = {2024},
	issn = {2664-5912},
	url = {https://elibrary.imf.org/openurl?genre=journal&issn=2664-5912&volume=2024&issue=001&cid=546859-com-dsp-crossref},
	doi = {10.5089/9798400273025.063},
	language = {en},
	number = {001},
	journal = {Fintech Notes},
	author = "Ricci, Luca and Ahokpossi, Calixte  and Belianska, Anna and Khandelwal, Khushboo and Lee, Sunwoo and Li, {Grace B.} and Mu, Yibin and Quayyum, {Saad N.} and Nunez, {Silvia G.} and Ree, {Jack J.} and {Rietti Souto}, Marcos and Simione, {Felix F.}",
	month = mar,
	year = {2024},
	pages = {1},
}

@techreport{oziliSurveyCentralBanka,
	title = "A Survey of Central Bank Digital Currency Adoption in {African} countries",
	author = {Ozili, Peterson K.},
	number = "118794", 
	institution = "MPRA",
	url = {https://mpra.ub.uni-muenchen.de/118794/}
}

@Article{kaur2024cbdc,
journal={Metamorphosis: A Journal of Management Research},
author={Harshdeep Kaur and Rajwinder Kaur and Monita Mago and Manjit Singh and Kanika Mehta},
title="Exploring Factors Affecting Central Bank Digital Currency Adoption: A Perspective from {Generation Z}",
year={2024},
month={December},
pages={126-141},
volume={23},
number={2},
doi={10.1177/09726225241286932},
url={https://ideas.repec.org/a/sae/metjou/v23y2024i2p126-141.html},
}

@misc{RwandaPublishesNew2024,
	title = "Rwanda Publishes New {CBDC} Feasibility Study",
	url = {https://digitalpoundfoundation.com/rwanda-publishes-new-cbdc-feasibility-study/},
	month = may,
	year = {2024},
	author = "Digital Pound Foundation"
}

@misc{cafonBankingOperations,
	author = {Consumer Advocacy Foundation of Nigeria},
	title = "Banking Operations and Innovations in {Nigeria}: A Case Study of Fintech Companies ({Opay})",
	howpublished = {\url{https://www.cafon.org.ng/banking-operations-and-innovations-in-nigeria-a-case-study-of-fintech-companies-opay/}},
	year = {2025},
	note = {[Accessed 05-02-2026]},
}

@misc{techdeskAfricasFintech,
	author = {Philia Mic-Julius},
	title = "{Africa}'s Fintech Boom in 2025: How Mobile Money and Digital Banks Are Reshaping the Continent",
	howpublished = {\url{https://techdesk.africa/2025/10/15/africas-fintech-boom-in-2025-how-mobile-money-and-digital-banks-are-reshaping-the-continent/}},
    publisher    = {TechDesk Africa},
	year = 2025,
	note = {[Accessed 05-02-2026]},
}

@misc{dmarketforcesAfricas205B,
	author = {Julius Alagbe},
	title = "Africa's \$205{B} Crypto Wave Reshapes Trade and Financial Access",
	howpublished = {\url{https://dmarketforces.com/africas-205b-crypto-wave-reshapes-trade-and-financial-access/}},
	year = {2026},
	month = jan, 
	note = {[Accessed 05-02-2026]},
}

@misc{do4africaCryptocurrencyAdoption,
	author = {DO4Africa},
	title = "Cryptocurrency adoption in {Africa}",
	url = {https://www.do4africa.org/en/cryptocurrency-adoption-in-africa/},
	year = {},
	note = "[Accessed 05-02-2026]"
}

@misc{bitcoinUkraineRussia,
	author = {Kevin Helms},
	title = "{Ukraine, Russia, South Africa, Nigeria} among top countries by cryptocurrency adoption", 
	url = {https://news.bitcoin.com/ukraine-russia-south-africa-nigeria-cryptocurrency-adoption/},
	year = {2020},
	note = {[Accessed 05-02-2026]},
}

@InProceedings{Tsuchiya:WWW24,
        author = "Taro Tsuchiya and Alejandro Cuevas and Nicolas Christin", 
        title = "Identifying risky vendors in cryptocurrency {P2P} marketplaces", 
        booktitle = "Proceedings of the {33rd Web Conference (WWW'24)}", 
        year = 2024,
        month = may,
        pages = "99--110",
        address = "Singapore"
}

@misc{ProtestsCashShortage,
	title = "Protests over cash shortage as {Nigeria} banknote switch looms", 
	author = "Chinedu Asadu",
	howpublished = "AP News",
	url = {https://apnews.com/article/nigeria-government-africa-business-dd07ec9ac8d8f5b786347b64f5fa7a1f}
}

@techreport{onuegbu2025communicationawarenessacceptancedigital,
      title="Communication, Awareness and Acceptance of Digital Banking Amidst Cash Crunch in {Southeast} and {South-South}, {Nigeria}", 
      author={Okechukwu Christopher Onuegbu and Bettina Oboakore Agbamu and Belinda Uju Anyakoha and Ogonna Wilson Anunike},
      year={2025},
      month = apr,
      number={2504.10546 [econ.GN]},
      institution={arXiv},
      url={https://arxiv.org/abs/2504.10546},
      doi ={10.48550/arXiv.2504.10546}
}

@article{hamrick2021examination,
  title={An examination of the cryptocurrency pump-and-dump ecosystem},
  author={Hamrick, JT and Rouhi, Farhang and Mukherjee, Arghya and Feder, Amir and Gandal, Neil and Moore, Tyler and Vasek, Marie},
  journal={Information Processing \& Management},
  volume={58},
  number={4},
  pages={102506},
  year={2021},
  publisher={Elsevier},
  url={https://doi.org/10.1016/j.ipm.2021.102506}
}

@inproceedings{xu2019anatomy,
  title="The anatomy of a cryptocurrency Pump-and-Dump scheme",
  author={Xu, Jiahua and Livshits, Benjamin},
  booktitle="Proc. 28th USENIX Security Symposium (USENIX Security 2019)",
  pages={1609--1625},
  year={2019}
}

@techreport{momo2025,
    author = {Raithatha, Rishi and Storchi, Gianluca},
    title = {The State of the Industry Report on Mobile Money 2025},
    institution = {GSMA},
    year = {2025},
    url = {https://www.gsma.com/sotir/wp-content/uploads/2025/04/The-State-of-the-Industry-Report-2025_English.pdf}
}

@inproceedings{olaleye2017users,
  title="Users experience of mobile money in {Nigeria}",
  author={Olaleye, Sunday A. and Sanusi, Ismaila T. and Oyelere, Solomon S.},
  booktitle={Proc. 2017 IEEE AFRICON},
  pages={929--934},
  year={2017},
  location={Cape Town, South Africa},
  publisher = {IEEE},
  address = {New York, NY, USA},
  doi = {10.1109/AFRCON.2017.8095606}
}

@article{adaramola2025dark,
  title={The dark arts of crypto laundering and the Nigerian financial ecosystem: Examining regulatory perspectives of virtual assets and virtual asset providers},
  author={Adaramola, Oluwabunmi},
  journal={Journal of Economic Criminology},
  volume={7},
  pages={100117},
  year={2025},
  doi = {10.1016/j.jeconc.2024.100117}
}

@techreport{chainalysis2024,
  title={The 2024 Geography of Crypto Report},
  author="Chainalysis",
  year={2024},
  month = oct, 
  institution={Chainalysis},
  address={New York}
}

@misc{cornell2025zig,
  title="What is {ZIG}, {Zimbabwe}’s Gold-Backed Digital Token?", 
  author="Chizunza, Jacob and  Chimhofu, Nigel Albert",
  year=2025,
  month = mar, 
  howpublished={Cornell SC Johnson College of Business Blog Post},
  url={https://business.cornell.edu/article/2025/03/what-is-zig/}
}

@techreport{davidwest2023cbdc,
  title="{CBDC} field research insights: Nigeria'sa {eNaira} -- Enabling possibilities",
  author="David-West, Olayinka and Umukoro, Immanuel",
  year=2023,
  institution={UC Irvine Institute for Money, Technology and Financial Inclusion},
  url={https://escholarship.org/uc/item/5kh8h2t5}
}

@incollection{davidwest2017agent,
  title="Adoption and use of mobile money services in Nigeria", 
  pages = "2724--2738", 
  doi = {10.4018/978-1-5225-2255-3.ch237}, 
  author="David-West, Olayinka and Umukoro, Immanuel Ovemeso and Muritala, Omotayo",
  booktitle={Encyclopaedia of Information Science and Technology, Fourth Edition},
  year={2017},
  publisher={IGI Global},
}

@techreport{ecb2024consumer,
  title="Consumer attitudes towards a central bank digital currency",
  author="Dimitris Georgarakos and Geoff Kenny and Luc Laeven and Justus Meyer",
  year=2025,
  institution="European Central Bank",
  type={ECB Working Paper Series},
  number={3035},
  url={https://www.ecb.europa.eu/pub/pdf/scpwps/ecb~cde4bd616e.wp3035en.pdf}
}

@techreport{imf2024cbdc,
  title={Central bank digital currency adoption: Inclusive strategies for intermediaries and users},
  author={{International Monetary Fund}},
  year={2024},
  type={FinTech Notes},
  number={2024/005},
  institution={International Monetary Fund}
}

@article{jalan2023trust,
  title={The role of interpersonal trust in cryptocurrency adoption},
  author ="Akanksha Jalan and Roman Matkovskyy and Andrew Urquhart and Larisa Yarovaya",
  journal={Journal of International Financial Markets, Institutions \& Money},
  volume={83},
  pages={101715},
  year={2023},
  month = mar,
  doi = {https://doi.org/10.1016/j.intfin.2022.101715}
}

@article{kumar2024drivers,
  title="Drivers influencing the adoption of cryptocurrency: A social network analysis approach",
  author="K. Kajol and Srijanani Devarakonda and Ranjit Singh and H. Kent Baker",
  journal={Financial Innovation},
  volume={11},
  number={74},
  year={2025},
  doi={10.1186/s40854-025-00757-0}
}

@article{liu2025cbdc,
  title={Antecedents of consumers' acceptance of central bank digital currency: The role of technology perceptions, social influence and personal traits},
  author={Liu, Xin and Wu, Jiaqi and Zhang, Chenghu},
  journal={Technological Forecasting and Social Change},
  volume={217},
  pages={124192},
  year={2025},
  doi={10.1016/j.techfore.2025.124192}
}

@article{nguyen2025cryptocurrency,
  title={Understanding cryptocurrency adoption: The role of technology, users, and trust in unregulated markets},
  author="Tian Le Nguyen and Van Kien Pham and Thi Thuy Dung Pham",
  journal={Human Behavior and Emerging Technologies},
  volume={2025},
  pages={7750468},
  year={2025},
  doi = {10.1155/hbe2/7750468}
}

@article{shahzad2024cryptocurrency,
  title={Cryptocurrency awareness, acceptance, and adoption: The role of trust as a cornerstone},
  author = "Muhammad Farrukh Shahzad and Shuo Xu and Weng Marc Lim and Muhammad Faisal Hasnain",
  journal={Humanities and Social Sciences Communications},
  month =dec, 
  volume={11},
  number={1},
  pages={1--14},
  year={2024},
  doi = {10.1057/s41599-023-02528-7}
}

@article{sharma2023empirical,
  title= "An empirical study of user adoption of cryptocurrency using blockchain technology: Analysing role of success factors like technology awareness and financial literacy",
  author = "Vandana Kumari and Pradip Kumar Bala and Shibashish Chakraborty", 
  journal={Journal of Theoretical and Applied Electronic Commerce Research},
  volume=18,
  number = 3,
  pages = "1580--1600",
  year = 2023,
  doi = {10.3390/jtaer18030080}
}

@techreport{tgm2024nigeria,
  title={Nigeria Crypto Insights Report 2024},
  author="{TGM Research}",
  year={2024},
  institution={TGM Research},
  url = {https://tgmresearch.com/nigeria-crypto-insights-2024.html}
}

@techreport{wezel2023nigeria,
  title={Nigeria: Fostering financial inclusion through digital financial services},
  author={Wezel, T. and Ree, J. K.},
  year={2023},
  type={IMF Selected Issues Paper},
  number={SIP/2023/020},
  institution={International Monetary Fund}
}

@article{tandon2022know,
  title="I know what you did on {Venmo}: Discovering privacy leaks in mobile social payments",
  author={Tandon, Rajat and Charnsethikul, Pithayuth and Arora, Ishank and Murthy, Dhiraj and Mirkovic, Jelena},
  journal={Proceedings on Privacy Enhancing Technologies},
  volume={3},
  pages={200--221},
  doi={10.56553/popets-2022-0069},
  year={2022},
  publisher={Privacy Enhancing Technologies Symposium},
  address={}
}

@misc{chainalysis2025subsaharan,
  title="{Sub-Saharan Africa} Emerges as Third-Fastest Growing Crypto Region with Strong Retail Activity",
  author="Chainalysis",
  year=2025,
  month = sep, 
  howpublished={Chainalysis Blog},
  url={https://www.chainalysis.com/blog/subsaharan-africa-crypto-adoption-2025/}
}

@misc{coingeek2024stablecoins,
  title="Stablecoins make up 43 percent of {Africa} crypto transactions in 2024",
  author="Steve Kaaru", 
  year={2025},
  month=sep, 
  howpublished={CoinGeek},
  url={https://coingeek.com/stablecoins-make-up-43-percent-of-africa-crypto-transactions-in-2024/}
}

@article{akomolehin2025cryptocurrencies,
  title={Cryptocurrencies as an Inflation Hedge: A Comparative Study Across High-Inflation Economies},
  author={Akomolehin F. Olugbenga},
  journal={International Journal of Research and Innovation in Social Science (IJRISS)},
  volume={9},
  number={10},
  pages={8526--8545},
  year={2025},
  month=nov,
  doi={10.47772/IJRISS.2025.910000694},
  issn={2454-6186}
}

@article{Garita2024Stablecoins,
  author       = {Garita, Mauricio and Cerezo Bregni, Celso Fernando and Asturias, Rodrigo},
  title        = "Stablecoins and inflation in {Latin America}: the case of {Argentina}",
  journal      = {Journal of Strategy and Management},
  year         = {2024},
  doi          = {10.1108/JSMA-05-2023-0119},
}

@misc{einisCashScarceZimbabwe2023,
	title = {Cash Is Scarce in {Zimbabwe}, As Inflation Spirals},
	url = {https://www.paymentsjournal.com/cash-is-scarce-in-zimbabwe-as-inflation-spirals/},
	urldate = {2026-02-06},
	howpublished = {PaymentsJournal},
	author = {Einis, Josh},
	month = mar,
	year = 2023
}

@misc{CashShortageHits,
	author="AfricaNews",
    title = "Cash shortage hits {Zimbabwe} banks as thousands stranded for {Christmas}",
	howpublished = "Africanews",
	url = {https://www.africanews.com/2021/12/23/cash-shortage-hits-zimbabwe-banks-as-thousands-stranded-for-christmas/},
	urldate = {2026-02-06},
    month={August},
    year={2024}
}

@article{McDonald:CSCW19,
  author = {McDonald, Nora and Schoenebeck, Sarita and Forte, Andrea},
  title = {Reliability and Inter-rater Reliability in Qualitative Research: Norms and Guidelines for CSCW and HCI Practice},
   volume={3},
  number={CSCW},
  articleno={72},
  year = 2019,
  month = nov,
  url = {https://doi.org/10.1145/3359174},
  doi = {10.1145/3359174},
  journal = {Proc. ACM CSCW 2019}
}

@article{Venkatesh:MISQ03,
 ISSN = {02767783},
 author = {Viswanath Venkatesh and Michael G. Morris and Gordon B. Davis and Fred D. Davis},
 journal = {MIS Quarterly},
 number = {3},
 pages = {425--478},
 publisher = {Management Information Systems Research Center, University of Minnesota},
 title = {User Acceptance of Information Technology: Toward a Unified View},
 urldate = {2026-02-05},
 volume = {27},
 year = {2003}
}

@article{qiu2019ripple,
  title="{Ripple vs. SWIFT:} Transforming cross border remittance using blockchain technology",
  author={Qiu, Tianyi and Zhang, Ruidong and Gao, Yuan},
  journal={Procedia computer science},
  volume={147},
  pages={428--434},
  year={2019},
  publisher={Elsevier},
  doi = {10.1016/j.procs.2019.01.260}
}

@misc{SI2020080Banking,
    author={Minister of Finance, Economic Development and Investment Promotion in Zimbabwe},
    title = {Banking (Money Transmission, Mobile Banking and Mobile Money Interoperability) Regulations, 2020},
	url = {https://www.veritaszim.net/node/4043},
	urldate = {2026-03-11},
    year={2020}
}

@article{davis1989technology,
  title={Technology acceptance model},
  author={Davis, Fred D and Bagozzi, Richard P and Warshaw, Paul R},
  journal={Journal of Management Science},
  volume={35},
  number={8},
  pages={982--1003},
  year={1989},
  publisher={Springer}
}

@article{venkatesh2003user,
  title={User acceptance of information technology: Toward a unified view1},
  author={Venkatesh, Viswanath and Morris, Michael G and Davis, Gordon B and Davis, Fred D},
  journal={MIS quarterly},
  volume={27},
  number={3},
  pages={425--478},
  year={2003},
  publisher={Management Information Systems Research Center, University of Minnesota}
}

@inproceedings{Tronnier2024,
author = {Tronnier, Frédéric},
year = {2024},
month = {01},
pages = {},
title = {Using Contextual Integrity to Uncover Acceptability of Information Flows in Central Bank Digital Currency Transactions},
doi = {10.24251/HICSS.2024.708}
}

@inproceedings{Pal2018,
author = {Pal, Joyojeet and Chandra, Priyank and Kameswaran, Vaishnav and Parameshwar, Aakanksha and Joshi, Sneha and Johri, Aditya},
year = {2018},
month = {04},
pages = {1-13},
title = {Digital Payment and Its Discontents: Street Shops and the Indian Government's Push for Cashless Transactions},
doi = {10.1145/3173574.3173803}
}

@inproceedings{Lewis2019,
author = {Lewis, Makayla and Perry, Mark},
year = {2019},
month = {04},
pages = {1-14},
title = {Follow the Money: Managing Personal Finance Digitally},
isbn = {978-1-4503-5970-2},
doi = {10.1145/3290605.3300620}
}

@article{he2023have,
  title={" I Have to Use My Son's QR Code to Run the Business": Unpacking Senior Street Vendors' Challenges in Mobile Money Collection in China},
  author={He, Changyang and He, Lu and Lu, Zhicong and Li, Bo},
  journal={Proceedings of the ACM on Human-Computer Interaction},
  volume={7},
  number={CSCW1},
  pages={1--28},
  year={2023},
  publisher={ACM New York, NY, USA}
}

@article{zhou2013empirical,
  title={An empirical examination of continuance intention of mobile payment services},
  author={Zhou, Tao},
  journal={Decision support systems},
  volume={54},
  number={2},
  pages={1085--1091},
  year={2013},
  publisher={Elsevier}
}

@article{chen2017understanding,
  title={Understanding continuance intention of mobile payment services: an empirical study},
  author={Chen, Xiaogang and Li, Shaorui},
  journal={Journal of Computer Information Systems},
  volume={57},
  number={4},
  pages={287--298},
  year={2017},
  publisher={Taylor \& Francis}
}

@article{odoom2020mobile,
  title={Mobile money usage and continuance intention among micro enterprises in an emerging market--the mediating role of agent credibility},
  author={Odoom, Raphael and Kosiba, John Paul},
  journal={Journal of Systems and Information Technology},
  volume={22},
  number={1},
  pages={97--117},
  year={2020},
  publisher={Emerald Publishing Limited}
}

@article{alhassan2020consumer,
  title={Consumer acceptance and continuance of mobile money: Secondary data insights from Africa using the technology acceptance model},
  author={Alhassan, Abidin and Li, Leon and Reddy, Krishna and Duppati, Geeta and others},
  journal={Australasian Journal of Information Systems},
  volume={24},
  year={2020},
  publisher={Australasian Association for Information Systems}
}

@article{al2024people,
  title={Why do people choose to continue using cryptocurrencies?},
  author={Al-Omoush, Khaled Saleh and Gomez-Olmedo, Ana M and Funes, Andr{\'e}s G{\'o}mez},
  journal={Technological Forecasting and Social Change},
  volume={200},
  pages={123151},
  year={2024},
  publisher={Elsevier}
}

@article{he2025antecedents,
  title={Antecedents of the continuance intention to use CBDCs: Evidence from the e-CNY},
  author={He, Liang and Jiang, Ruiqi and Wang, Chao and Yao, Yue},
  journal={International Review of Economics \& Finance},
  pages={104870},
  year={2025},
  publisher={Elsevier}
}

@article{perry2018moneywork,
  title={Moneywork: Practices of use and social interaction around digital and analog money},
  author={Perry, Mark and Ferreira, Jennifer},
  journal={ACM Transactions on Computer-Human Interaction (TOCHI)},
  volume={24},
  number={6},
  pages={1--32},
  year={2018},
  publisher={ACM New York, NY, USA}
}

@article{chandra2010evaluating,
  title={Evaluating the role of trust in consumer adoption of mobile payment systems: An empirical analysis},
  author={Chandra, Shalini and Srivastava, Shirish C and Theng, Yin-Leng},
  journal={Communications of the association for information systems},
  volume={27},
  number={1},
  pages={29},
  year={2010}
}

@article{xin2015antecedents,
  title={Antecedents of consumer trust in mobile payment adoption},
  author={Xin, Hua and Techatassanasoontorn, Angsana A and Tan, Felix B},
  journal={Journal of Computer Information Systems},
  volume={55},
  number={4},
  pages={1--10},
  year={2015},
  publisher={Taylor \& Francis}
}
